\documentclass[twocolumn,3p,10pt]{elsarticle}
\usepackage{lineno,hyperref}
\usepackage[numbers]{natbib}
\usepackage[utf8]{inputenc}
\usepackage{amsthm}
\usepackage{multirow}
\usepackage{booktabs}
\journal{Journal of \LaTeX\ Templates}

\begin{document}

\begin{frontmatter}

\title{First-principles based Monte Carlo modeling of oxygen deficient Fe-substituted SrTiO$_{3}$ experimental magnetization}

\author[mycorrespondingauthor]{Juan M. Florez}
\ead{caross@mit.edu,juanmanuel.florez@usm.cl}
\author[mycorrespondingauthor]{Miguel A. Solis}
\author[mycorrespondingauthor]{Emilio A. Cort\'{e}s Estay}
\author[mycorrespondingauthor,mymainaddress]{E. Su\'arez Morell}

\address[mycorrespondingauthor]{Grupo de Simulaciones, Departamento de F\'{i}sica, Universidad T\'ecnica Federico Santa Mar\'{\i}a, Valpara\'{i}so 2390123, Chile\\}
\address[mymainaddress]{Departamento de Física Aplicada I, Escuela Politécnica Superior, Universidad de Sevilla, España}

\author[mysecondaryaddress]{Caroline A. Ross}
\address[mysecondaryaddress]{Department of Materials Science and Engineering, Massachusetts Institute of Technology, Cambridge, MA 02139, USA}
%\author[]{Elsevier Inc}

%\ead{}

%+++++++++++++++++++++++++++++++++++++++++++++++++++++++++++++++++++++++++
%+++++++++++++++++++++++++++++++++++++++++++++++++++++++++++++++++++++++++
\begin{abstract}
Ferroics based on transition-metal (TM) substituted SrTiO$_{3}$ have called much attention as magnetism and/or ferroelectricity can be tuned by using cations substitution and defects, strain and/or oxygen deficiency. C. A. Ross et al. [Phys. Rev. Applied 7, 024006 (2017)] demonstrated the SrTi$_{1-x}$Fe$_{x}$O$_{3-\delta}$ (STF) magnetization behavior for different deposition oxygen-pressures, substrates and magnetic fields. The relation between oxygen deficiency and ferroic orders is yet to be well understood, for which the full potential of oxygen-stoichiometry engineered materials remain an open question. Here, we use hybrid-DFT to calculate different oxygen vacancy ($v_{o}$) states in STF with a variety of TM distributions. The resulting cations' magnetic states and alignments associated to the $v_{o}$ ground-states for $x=\{0.125,0.25\}$ are used within a Monte Carlo scope for collinear magnetism to simulate the spontaneous magnetization. Our model captures several experimental STF features i.e., display a maximum of the magnetization at intermediate number of vacancies, a monotonous quenching from $\sim{0.35}\mu{_{B}}$ for small ${\delta}$, and a slower decreasing of such saturation for larger number of vacancies. Moreover, our approach gives a further insight into the relations between defects stabilization and magnetization, vacancy density and the oxygen pressure required to maximize such ferroic order, and sets guidelines for future Machine Learning based computational synthesis of multiferroic oxides.
\end{abstract}
\begin{keyword}
magnetic perovskites, Monte Carlo, oxygen deficiency, density functional theory
%\texttt{elsarticle.cls}\sep \LaTeX\sep Elsevier \sep template
%\MSC[2010] 00-01\sep  99-00
\end{keyword}
\end{frontmatter}
%\linenumbers
%+++++++++++++++++++++++++++++++++++++++++++++++++++++++++++++++++++++++++
%+++++++++++++++++++++++++++++++++++++++++++++++++++++++++++++++++++++++++
\section{Introduction}
Perovskite-structured oxides exhibit an exceptionally rich variety of electronic properties including ferroelectricity \cite{Ferrooxide_superla,BiFeO_multi_hetero,BaTiOFerroel_film,Ferroelec_strin_sto}, magnetic order, superconductivity \cite{superiso_sto,supersemi_sto,superinter_bet_oxi,nonstoi_grain_sto}, and multiferroicity \cite{multiferroico1,OdefSTF,multiferro_roomt_ortho-morpho_lfo}, properties that can be tuned via the composition, doping, strain state, and defect population of the material\cite{OdefSTF,Ferroindu_isoto_exchan_sto,anti-ferrodist_vacan_sto,oxy_vac_stco_teo-exp,pointdefect_ferroelec_sto,large_magne_Odefici_sto,electro_doping_metal_oxide,Sikam:2018daa}. One of the cornerstones of oxide electronics is SrTiO$_{3}$ (STO) which has a large bandgap and nonmagnetic behavior at room temperature\cite{sto_with_hse}. STO can be integrated with Si devices, and ferroelectric and magnetic behavior can be promoted by strain and by magnetic substitution, respectively \cite{multiferroico1,OdefSTF,Sikam:2018daa,sto_with_hse,Pai:2018fia}. Recent work has focused on the  role of oxygen-defects and introduction of Ti-substituent on the multiferroic properties \cite{OdefSTF,anti-ferrodist_vacan_sto,oxy_vac_stco_teo-exp,pointdefect_ferroelec_sto,large_magne_Odefici_sto,electro_doping_metal_oxide,Sikam:2018daa,Pai:2018fia,MagSTOdelta,OrbitalSymmetrySTO,StrainControl,Dong:2018dr,Lee:2018dka,Schiaffino:2017cwa,Brovko:2017hh,Wang:2017ii}, and an interplay between the oxygen stoichiometry and the magnetic and ferroelectric degrees of freedom has been demonstrated.\\

Stoichiometry in ABO$_{3}$ perovskites was thought to be a key to obtain robust magnetization \cite{gerald dionne book}, and the introduction of oxygen vacancies ($v_{O}$) or cation defects usually led to weak magnetic ordering or paramagnetic-like states \cite{OdefSTF,STFexp}. However, experimental results have shown that O-deficiency is capable of turning a magnetic semimetal such as SrCoO$_{3}$ (SCO) into a semiconductor as well as converting the insulating paramagnetic STO into a magnetic semiconductor or weak ferroelectric \cite{multiferroico1,Pai:2018fia,STC}. $3d$-orbitals in transition metal (TM) cations are distorted by incomplete oxygen octahedral coordination O$_{5,4}$, and both the covalent and ionic bondings characters through O-A/B-$v_{O}$ defects plays a relevant role \cite{oxidation_partialcharge_ionicity}, as in the case of the multi-interpreted SCO spin-states \cite{STC} and the magnetism in Fe,Co-substituted STO\cite{STFC_Ox_hybrid,STFC_polar}.\\

On the other hand, ferroic ordering can be engineered through defects in ABO$_{3}$ perovskites, e.g. at low temperatures, stoichiometric STO presents antiferrodistortive structural changes and quantum fluctuations that suppress the ferroelectric (FE) ordering \cite{STO_Review,STO_comp_AFD-FE}. Also, coupling of interstitial and anti-site Ti with Sr and O vacancies, have been suggested to promote polar effects beside magnetism \cite{STO_SrOO_vac,STO_Ti_antisite,STO_Pol_defects,STO_FE_SrTi_ratio,STO111,STO001}. Multiferroism in oxygen deficient SrFeO$_{3}$ nanoparticles was studied \cite{ferro6} and it displayed saturation electric-polarization depending on the Fe concentration \cite{Brovko:2017hh}. Moreover, electric polarization was realized in magnetic Fe-doped Ti-rich STO at room temperature\cite{STF_multiferroic2}. Among TM-substituted STO, SrTi$_{1-x}$Fe$_x$O$_{3-\delta}$ (STF) and SrTi$_{1-x}$Co$_{x}$O$_{3-\delta}$ (STC) both display magnetization that depends on their oxygen content, with typically higher magnetization at higher levels of oxygen deficiency ($\delta$), and a distinct magnetization maxima characterizing STF \cite{OdefSTF, STC, STC2018}.\\

In this work, we focus on the magnetic properties of STF. Its magnetic properties can be varied via both cation composition and O deficiency. Moreover, room temperature magnetism and anisotropy with out-of-plane magnetic easy axis are observed in thin films of nondilute STF and STC deposited on different substrates, showing that both strain and oxygen deficiency are key factors in determining the magnetic properties of substituted STO, and the effect of oxygen pressure during growth has also been explored \cite{OdefSTF,STC}. The magnetization of STC and STF increases  with the Co or Fe  content, and is higher for lower growth pressure and therefore greater oxygen vacancy concentration, although for STF a decrease in magnetization was found at the lowest growth pressures \cite{OdefSTF,STFexp,STC,roomt1,roomt2,FMSTCdelta}. Using XMCD we showed that the magnetic moment in STF was proportional to the concentration of divalent Fe which increased on films grown at low pressures or annealed in a reducing environment \cite{OdefSTF,STFexp}. The magnetic properties are intrinsically tied to the allowed mixture of valence and spin states and the corresponding ferromagnetic (FM) or antiferromagnetic (AFM) local spin-ordering, which strongly depends upon the oxygen vacancies coordinating the transition metal ions.\\

A comprehensive model of the perovskite vacancies distribution and its repercussion with respect to the ferroic order parameters remains a challenge. Moreover, in STF, experimental magnetization can not be completely interpreted as a direct consequence of the changes in the cations valence spin states and TM symmetry alone \cite{OdefSTF,STFexp}, for which the distinction between different vacancy stabilized local spin orderings and breaking of symmetry should be taken into account in a wider way next.
We theoretically investigate the spontaneous/saturation magnetization of STF. First, we use hybrid-DFT methods to obtain the STF spin-states, local magnetic ordering, and energetically favored TM symmetry and defects formation; then use that microscopic information to feed an intuitive statistical model for the magnetization observable, whose probability distribution is calculated by using a Monte Carlo method that samples over the vacancies configurations space for any given number of vacancies, while the symmetry and magnetic constrictions imposed by the ground state solutions conforming such configurations are fulfilled. 
STF magnetization maximization result for an intermediate oxygen pressure has not been addressed either partly because of the difficulty in terms of the large configurational space to be considered besides the variety of thermodynamic and  chemical-physicist conditions related to the deposition and synthesis. Our microscopic and statistical approaches both are to a good extent in agreement with STF experiments and might apply to other ABO$_{3}$ systems.\\

This article is organized as follows: In section 2 we introduce the perovskite model, the DFT methods, and the MC approach. In section 3 we show the DFT results for stoichiometric and oxygen deficient systems with $\delta=\{0,0.125\}$, and $x=\{0.125,0.25\}$. In section 4 we present the MC statistical analysis for the STF magnetization. In section 5 the conclusions are presented. 
%+++++++++++++++++++++++++++++++++++++++++++++++
\begin{figure*}[ht]
\center
\includegraphics[width=0.6\linewidth]{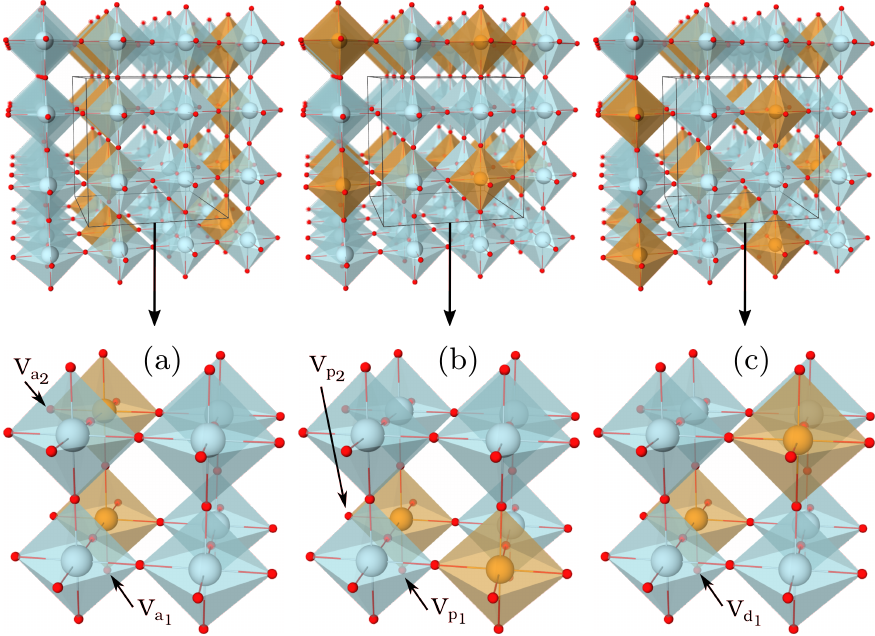}
\vspace{0.0in}
\caption{Supercell for $x=0.25$ in SrTi$_{1-x}$Fe$_{x}$O$_{3-\delta}$: (a) Fe-Fe pair with $nn_{\mathrm{Fe}}={a}$. (b) Fe-Fe pair with $nn_{\mathrm{Fe}}={a}\sqrt{2}$;
and (c) Fe-Fe pair with $nn_{\mathrm{Fe}}={a}\sqrt{3}$. $a$ is initially the STO lattice parameter. The vacancies are depicted as: $V_{a_{1,2}}$, $V_{p_{1,2}}$ 
and $V_{d_{1}}$, respectively. The Sr atoms have been omitted for the sake of simplicity. Graphical representations were produced using the OVITO software \cite{ovito}.
\vspace{0.0in}
\label{figure:models}}
\end{figure*}
%+++++++++++++++++++++++++++++++++++++++++++++++
%++++++++++++++++++++++++++++++++++++++++++++++++++++++++++++++++++++++++
%++++++++++++++++++++++++++++++++++++++++++++++++++++++++++++++++++++++++
\section{Modeling framework}
We use density functional theory with a Heyd-Scuseria-Ernzerhof (HSE06) exchange-correlation functional 
as implemented in the Vienna Abinitio Simulation Package (VASP 5.4) \cite{vasp96prb,vaspbackground,HSE061} to study the effects of oxygen deficiency in STF. The use of a hybrid functional leads to improved accuracy over standard local and semi-local functionals such as LSDA, GGA or Perdew-Burke-Ernzerhof (PBE) for predictions of key properties such as valence spin-states, chemical-induced structural changes and band-gaps in 3d-oxides as shown previously e.g., for SCO, STO, STC and SrTi$_{1-x-y}$Fe$_x$Co$_{_y}$O$_{3-\delta}$ (STFC) \cite{OdefSTF,STFC_Ox_hybrid,STC}. We also performed GGA+U calculations in order to examine the stability of selected Fe spin-states under strain for the sake of discussion, because  HSE06 functional incurs in large computational costs compared to local methods.\\ 

The spin-polarized calculations were performed with an energy cutoff of 500 eV for $2\times2\times2$ and $4\times4\times4$ k-point grids in the case of HSE06, for relaxations and static calculations, respectively, and $6\times6\times6$ for GGA+U in both cases. The HSE06 grids were chosen to keep the computational costs at a reasonable level, and the results represent converged relaxations with forces below {$10^{-4}$ eV/\AA}. Our screening value $\mu$ was  chosen similar to References \cite{OdefSTF,STC} to  compare with STC/STO results, and the $U_{eff} = U - J$ terms were used within  Dudarev's GGA+U approach for the $d$-Fe electrons \cite{dudarev}. In the latter case, the  acting forces were reduced below {$10^{-5}$eV/\AA}, and the $U_{eff}$ values were chosen within the range of validity of recent simulations for STFC i.e., $4$ and $7$ eV for $3d$-orbitals were applied to Fe and Ti, respectively \cite{STFC_polar}. In the case of the ab-initio calculations, the supercell of the solid solutions consisted of $2\times2\times2$ unit cells with $\{40,39\}$ ions for $\delta=\{0.0,0.125\}$, as displayed in Figure \ref{figure:models}. PBE pseudopotentials were used with $3s^{2}$$4p^{6}$$5s^{2}$, $3d^{3}$$4s^{1}$, $3d^{7}$$4s^{1}$, and $2s^{2}$$2p^{4}$ distributions for Sr, Ti, Fe, and O respectively.\\
%+++++++++++++++++++++++++++++++++++++++++++++++
\begin{figure*}[t]
\center
%\vspace{0.0in}
\includegraphics[width=1.0\linewidth]{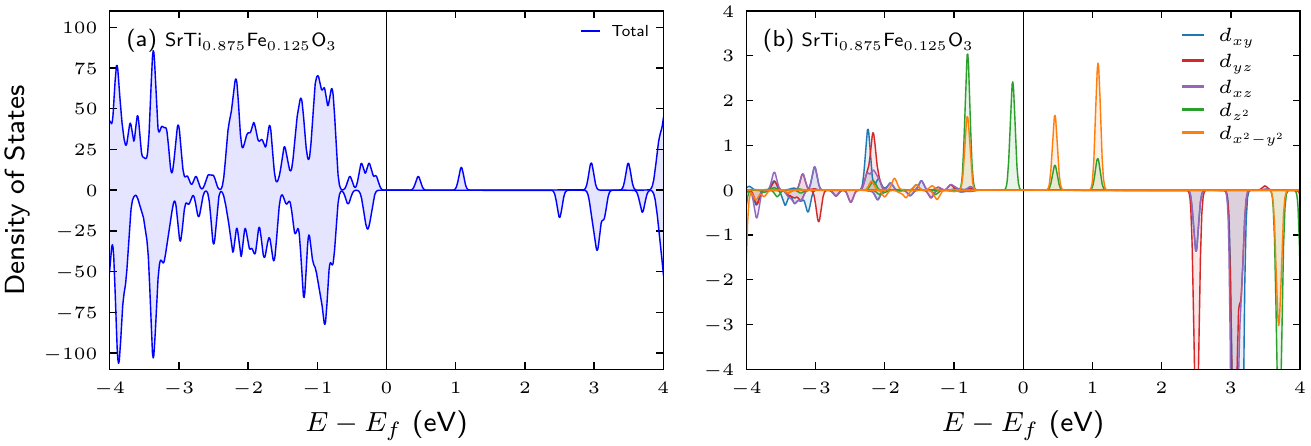}
%\vspace{0.0in}
\caption{(a) Total Density of States (DOS) and (b) Fe ion $d$-orbitals projected DOS for SrTi$_{0.875}$Fe$_{0.125}$O$_3$.
%\caption{(a) Total Density of States (DOS) for Sr$^{2+}$Ti$^{4+}_{0.875}$Fe$^{4+}_{0.125}$O$^{2-}_3$. (b) $d$-orbitals projected DOS for the same system (HSE06).
\label{figure:totalstoichio125}}
\end{figure*}
%+++++++++++++++++++++++++++++++++++++++++++++++

The positions of the oxygen vacancies, shown in lower panels of Figure \ref{figure:models} as $V_{{a,p,d}_{(1,2)}}$, were selected after a symmetry analysis of the relaxed stoichiometric systems corresponding to three different distributions of the  Fe cations within the supercells for $\%25$ Fe substitution,  obtained with FINDSYM and pymatgen applications \cite{pymatgen,matproj,FINDSYM} for a {$10^{-3}$\AA} tolerance. The Fe-nearest-neighbor (nn) distances considered were $\{a$,$\sqrt{2}a$,$\sqrt{3}a\}$, with $a$ the initial unit-cell parameter. According to each configuration, we have considered all the atomic valence states, which are reflected in several possible high and low Pauli states for the TM, as well as the possible combinations for the FM or AFM Fe-Fe exchange coupling. The valences of the cations are selected to maintain a neutral supercell for a given oxygen deficiency, and point/bulk-charging effects due to these defects are negligible as suggested by DFT calculations in STF and STFC \cite{STFC_Ox_hybrid,STFC_polar}. We also refer in this work, without loss of generality, to $V_{{a,p,d}_{(1,2)}}$ as the vacancies corresponding to the Fe cations aligned along the $[1,0,0]$ (above), $[1,1,0]$ (plane) and $[1,1,1]$ (diagonal) crystalline directions, respectively.\\

To model the magnetic order parameter we use Monte Carlo (MC) \cite{miguel2} calculations to extract $v_\mathrm{O}$-configurational probabilities for specific cations distributions and using $4\times4\times4$ supercells as displayed in the upper panels of Figure \ref{figure:models}. Within a Metropolis \cite{miguel1} scheme an algorithm with an aleatory-sampling acceptance $\mathrm{rand(0,1)}\le\exp{(-\Delta{E}/k_{B}T)}$ is implemented. Here the $\Delta E$ energies are obtained from ab-initio characterization, such that they discriminate the states corresponding to FM from the AFM ones, as well as the lowest-energy states (gs) from second to lowest solutions (ss), for a representative group of $v_{o}$ out of the all $V_{{a,p,d}_{(1,2)}}$ possibilities.\\

%+++++++++++++++++++++++++++++++++++++++++++++++
\begin{figure}[t]
\center
\includegraphics[width=1.0\linewidth]{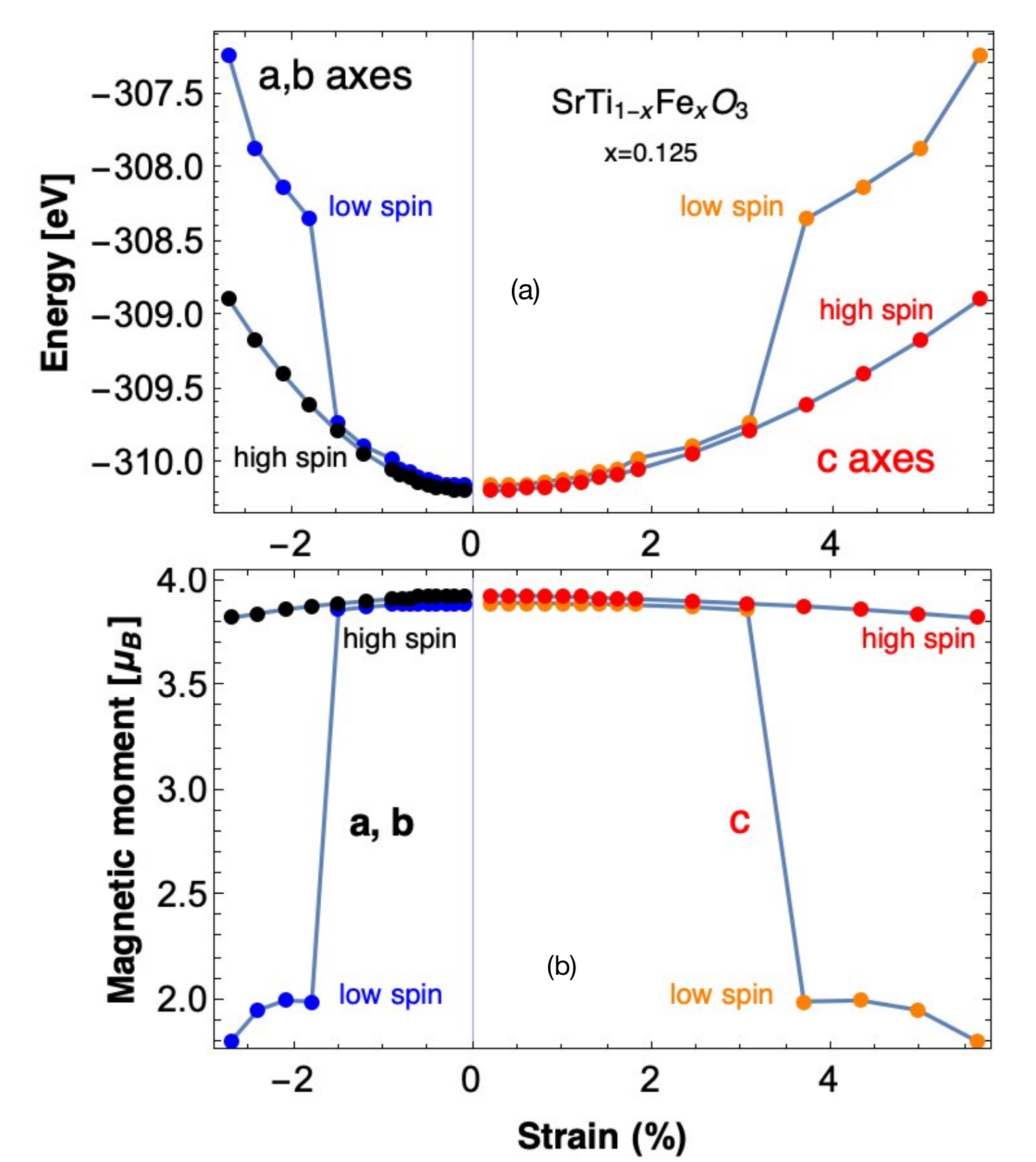}
\vspace{0.0in}
\caption{(a) Energy and (b) magnetic moment of STF with $x=0.125$ and $\delta = 0$, for several tetragonal strains: in relaxations (GGA+U) Fe are initialized with Fe$^{4+}$ low and high spin states. Black/blue curves correspond to the tetragonal $(a, b)$ axes while red/orange correspond to the out-of-plane $c$ axis.
\vspace{0.0in}
\label{figure:strainEneMag125}}
\end{figure}
%+++++++++++++++++++++++++++++++++++++++++++++++
Our MC simulation works by extracting oxygen ions from the lattice similar to how a Galton-board arranges balls \cite{miguel3}, where the resulting balls-filled columns' heights would represent the different $v_\mathrm{O}$-configurational probabilities for a certain number of vacancies arranged in the octahedra. As such, for a given adiabatic temperature and a number of $v_\mathrm{O}$, up to $1\times10^{5}$ MC trials are performed, including sets of trials for several MC random seeds, and the frequency of occurrence of a particular configuration characterized by a specific group of locations for the vacancies and with a specific local magnetic ordering is translated to the probability $P_{k}$, which is normalized over all the possible arrangements such that $\sum_{k}{P_{k}=1}$. The average magnetization $\langle M \rangle_{k}$ of each configuration ${k}$ is addressed by using the Fe magnetic moments predicted by the hybrid-DFT model for each $V_{{a,p,d}_{(1,2)}}$ type within the arrangement. The final magnetization for the deficient perovskite, for a given oxygen deficiency, is then calculated as:
%+++++++++++++++++++++++++++++++++++++++++++++++
\begin{equation}
\begin{array}{ccccc}    
 & & \langle M\rangle & = & \sum_{k}^{}{P_{k}} {\langle M \rangle}_{k}
\end{array}
\label{equa:proba}
\end{equation}
%+++++++++++++++++++++++++++++++++++++++++++++++

Equation \ref{equa:proba} uses a distribution that is obtained by sampling according to a stationary Markov process, i.e., the occurrence of a configuration with a certain number of vacancies depends on the probability of occurrence of a particular configuration with one less vacancy. \\

On the other hand, magnetic orderings used in the MC simulations are bound to the spin-polarization and supercell $v_\mathrm{O}$-density modeled through DFT here, which in this case means scenarios where magnetism is dominated by the cations interaction in collinear solutions, one $v_\mathrm{O}$/Fe-Fe pair approximately, as well as bath-temperatures for $k_b T$ below formation energies and spin-gaps (energy difference between FM to AFM local ordering). More aspects of our MC modeling are further discussed in next sections as well as in the appendices section. However, we will show next that these aforementioned approximations provide a powerful representation of one of the most important properties of a perovskite intrinsic magnet i.e., its spontaneous/saturation magnetization, and allows to understand better the role of the $v_\mathrm{O}$-density in magnetic oxides synthesis. \\

Finally, we used the Interface Reaction of The Materials Project \cite{materialsproject2} to examine the magnetism of the compounds resulting in the reaction of two STO-based perovskites in contact \cite{materialsproject3}. The oxygen partial pressure and/or possible temperature effects are modeled by using a single chemical potential variable\cite{materialsproject2,materialsproject3}. The predictor provides us with the magnetic components' chemical fractions, which are then used to estimate the average magnetic moment \cite{materialsproject3} using the Materials Project database for the electronic properties of those components, and compare with our MC results and experiments. 

%+++++++++++++++++
\begin{table}[t]
\caption{Sr$^{2+}$Ti$^{4+}_{0.875}$Fe$^{4+}_{0.125}$O$^{2-}_3$ (HSE06) lattice parameter, Fe magnetic moment and energies corresponding high/low spin initializations.}
\begin{center}
\begin{tabular}{c c c c c}
%\hline
$a'$ (\AA) & $S_{i}$ & $\mu_B/\mathrm{Fe}$ & $E_{S_{i}}$ (meV/f.u.)\\
\hline 
        7.81 & h & 3.73  & $\bf 0.0_{gs}$\\    
            7.79 & l & 1.98  & $68.2$\\                                 
\hline
\end{tabular}
\end{center}
\label{table1}
\end{table}
%+++++++++++++++++
%++++++++++++++++++++++++++++++++++++++++++++++++++++++++++++++++++++++++
%++++++++++++++++++++++++++++++++++++++++++++++++++++++++++++++++++++++++
\section{DFT Results}
\subsection{Stoichiometric STF: SrTi$_{0.875}$Fe$_{0.125}$O$_{3}$}
In ``stoichiometric" STF ($\delta = 0$) the Fe ion would take a 4+ valence state to keep charge balance. This valence state is present e.g., in STF synthesized at high oxygen pressures $\>>$ 1 atm \cite{Tuller}. Table \ref{table1} displays the energy characterization of STF as stabilized in high and low valence spin-states for $x=0.125$. It can be seen that the high spin in Fe corresponding to a  $(t_{2g}^{3},e_{g}^{1})$ $3d$ occupation is preferred over the low $(t_{2g}^{4},e_{g}^{0})$ case. Metallic Fe/Co-based compounds are usually expected to be found in low spin states e.g., well known oxides such as magnetite Fe$_{3}$O$_{4}$ or SCO itself. However, in magnetite, Fe valence is strongly affected by electron exchange between the tetrahedral and octahedral sites increasing the Fe-spin to an intermediate value \cite{jmflorezmagnetite}, while in the case of O-deficient SCO, an intermediate Co spin is favored by Co-Vacancy-Co coupling through the AFM shared electrons \cite{STC}. STF has a $\sim{0.4}$ eV band-gap, as can be seen in Figure \ref{figure:totalstoichio125}, which is bounded by the Fe-e$_{g}^{1}$ acceptor-like state,  with such high spin preference is also deviating from the simplest molecular predictions. \\

This high spin seems robust with respect to tetragonal distortions, which are among the first factors lowering the crystal symmetry due to, e.g., substrate mismatch. As shown in Figure \ref{figure:strainEneMag125}, large strains are needed to promote $10Dq$ changes. This last figure shows a high-to-low spin transition at $\sim 1.5\%$ in-plane strain. However, in order to capture such magnetic changes the relaxations have been constrained such that the perovskite single-crystal volume is averagely unchanged as it occurs, e.g., in STC where the substrate-mediated strain is balanced with chemical strain due to Co incorporation \cite{roomt2}. Full relaxed strain always leads to similar high-versus-low initialized magnetic results or elastically worse strain values. The ineffectiveness of these distortions are in accordance with Moreno et. al.\cite{moreno}, which suggested that metal-ligands distances are not so decisive in both the intrinsic e$_{g}$ splitting and cubic field splitting $10Dq$. Bondings with deep 2s (free atom) orbitals, rather than shallow 2p ones, are essential to change orbitals occupations \cite{STFC_Ox_hybrid,moreno}.
%+++++++++++++++++
\begin{table*}[h]
\caption{Sr$^{2+}$Ti$^{4+}_{1-x}$Fe$^{4+}_x$O$^{2-}_3$ properties for $x=0.25$ (HSE06). }
\begin{center}
\begin{tabular}{c c c c c}
%\hline
\multicolumn{5}{c}{$E_{S_{i}}$ (meV/f.u.) for configurations in Fig. \ref{figure:models} with $S_{i}$} \\
\cline{1-5}
    $nn_{\mathrm{Fe}_i}/a$ & $E_{h_{AFM}}$  &   $E_{h_{FM}}$ & $E_{l_{AFM}}$ & $E_{l_{FM}}$  \\ \hline
    $1$& 32.5 & \bf 0.0$\bf_{a}$  & 33.6 & 201.0    \\
    $\sqrt{2}$& \bf 35.7$\bf_{b}$ & 37.5 & 158.1 & 157.3 \\
    $\sqrt{3}$& 37.2 & \bf 36.4$\bf_{c}$ & 36.3 & 167.6 \\ \hline
    \multicolumn{5}{c}{Lattice parameters and magnetic structure for \bf{a, b, c}} \\ \hline
    $S_{f}^{1,2}$ ($\mu_B$) & $(a', b', c')$ (\AA)   & $V$ (\AA$^3$)   & $nn_{\mathrm{Fe}_f}/a$ & $E_{bg}$ (eV) \\ \hline
    $(3.7,~3.7)_{\bf a}$ &  (7.80,7.80,7.79) & 474.66 & $\sim0.89$ & 0.00\\
    $(3.7,-3.7)_{\bf b}$ & (7.80,7.80,7.80) & 474.70 & $\sim0.9\sqrt{2}$ & 0.01\\
    $(3.7,~3.7)_{\bf c}$ & (7.80,7.80,7.80) &  474.69 & $\sim0.9\sqrt{3}$ & 0.16\\
\hline
\end{tabular}
\end{center}
\label{table2}
\end{table*}
%+++++++++++++++++
%++++++++++++++++++++++++++++++++++++++++++++++++++++++++++++++++++++++++
%++++++++++++++++++++++++++++++++++++++++++++++++++++++++++++++++++++++++
\subsection{Stoichiometric STF: SrTi$_{0.75}$Fe$_{0.25}$O$_{3}$}
We focus from this point onward on the composition corresponding to $x=0.25$, which within our model perovskite means two Fe/u.c., as Figure \ref{figure:models} shows. These Fe cations can be arranged into three different distributions, which we considered and evaluated at high and low spin states for the corresponding valences and ferromagnetic/antiferromagnetic local spin orderings. In Table \ref{table2} hybrid relaxation results are presented, where $E_{(h,l)_{_{FM,AFM}}}$ describes the energy difference of the specific state with respect to the global ground state among all. ${{nn_{\mathrm{Fe}}}_{i,f}}/a$ and $S_{i,f}$ are the proper initial and final Fe-Fe supercell distances and local magnetic moment, respectively. We select for each Fe distribution the lowest energy state being the global ground state corresponding to the ferromagnetic $E_{h_{_{FM}}}$ high-spins ordering. The first two major suggestions of Table \ref{table2}, which already differ qualitatively from what happens for instance in STC \cite{STC} are: the magnetic cations stabilized preferably at a first ${nn_{\mathrm{Fe}}}$ configuration respect to the B-B (ABO$_{3}$) possibilities, Figure \ref{figure:models}(a); high spin states dominate the stabilized magnetism, which follows previous section' conclusions.\\ 

In Table \ref{table2} the energies of the low spin states that are comparable to those of high spin states are due to self-consistent switching to high spins. Further constrictions to force those low states lead to higher energies as already shown by Figure \ref{figure:strainEneMag125} and other results in Table \ref{table2}, with energies over $\sim{150}$ meV in this last case. Comparatively, the intermediate distance between the Fe ions is the one allowing an AFM ordering, while in STC it stabilizes the ground state in a FM configuration. 
In this $\delta=0$ case, for the systems it is energetically less expensive to arrange an AFM ordering by switching one of the two spins than stabilizing larger ${nn_{\mathrm{Fe}}}$. It is observed that the $E_{h_{_{AFM}}}$ energy increases as ${{nn_{\mathrm{Fe}}}}/a$ increases while keeping AFM local moments, however, if the local moments are kept FM the system seems to experience a disconnection when Fe ions are pulled apart. They stabilize the same in whatever location they are placed in, not without a huge energetic cost from the ground state. The results in Table \ref{table2} therefore witness the different exchange-coupling nature that links the Fe ions in these three configurations i.e., the oxygen mediates magnetic interactions through super-exchange mechanisms that in this Fe-substituting case seem negligible beyond B-B first nearest-neighbors. \\

Our stoichiometric STF in Table \ref{table2} is then suggesting that if a real sample were not to be mono-crystalline but is conformed by at least a couple of domains due to effects such as e.g. strain, chemical pressure, annealing, or cations/anions defects, a mixture of FM and AFM local orderings would be stabilized giving rise to an average magnetization observable independent of the resulting magnetic ions sub-lattice. However, the energy weight of a $[1,0,0]$ Fe-cations crystalline symmetry and the magnetic switching of this last one would be statistically preponderant among other configurations.
%+++++++++++++++++++++++++++++++++++++++++++++++
\begin{figure*}[t]
\center
\vspace{0.0in}
\includegraphics[width=0.99\linewidth]{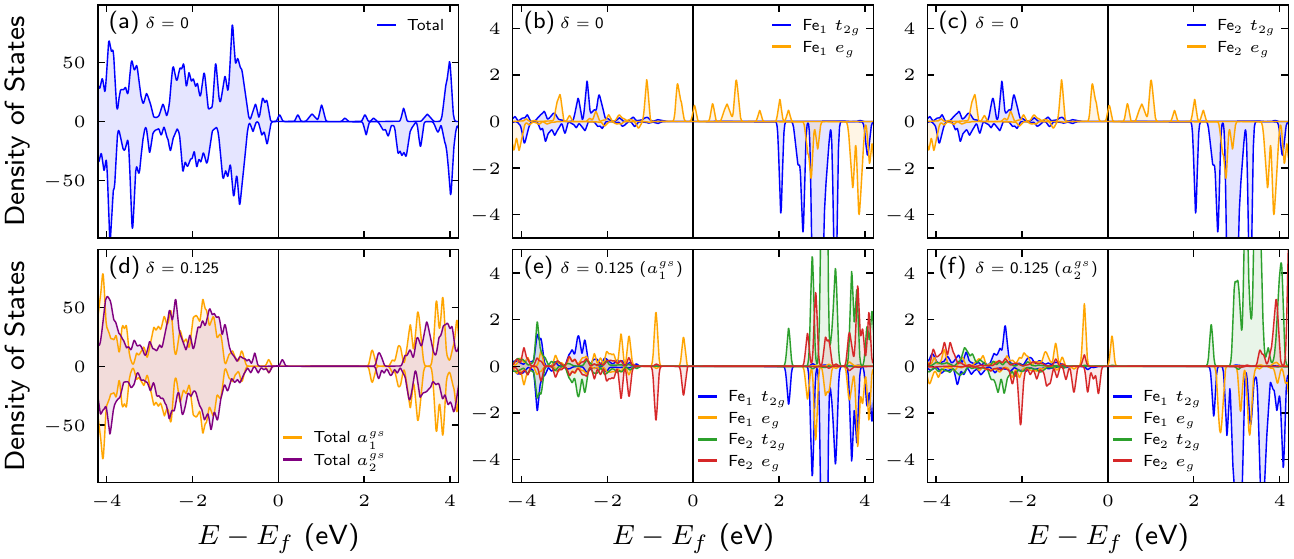}
\vspace{0.0in}
\caption{(a) Total DOS and (b, c) t$_{2g}$e$_{g}$-projected DOS of Fe$_{1}$ and Fe$_{2}$ for the ground state stoichiometric STF.
(d) Total DOS and (e, f) t$_{2g}$e$_{g}$-projected DOS of Fe ions for the V$_{a_{1,2}}$ vacancies ground states in SrTi$_{0.75}$Fe$_{0.25}$O$_{2.875}$.
\label{figure: total stoi a1 and a2 and t2g2g a1 and a2}}
\end{figure*}
%+++++++++++++++++++++++++++++++++++++++++++++++

%++++++++++++++++++++++++++++++++++++++++++++++++++++++++++++++++++++++++
%++++++++++++++++++++++++++++++++++++++++++++++++++++++++++++++++++++++++
\subsection{Oxygen deficient STF: SrTi$_{0.75}$Fe$_{0.25}$O$_{2.875}$}
For STF, the magnetization as a function of the oxygen pressure during growth was depicted in References \cite{OdefSTF,STFexp}, and it was clear that there is a tuning process that is triggered by the oxygen vacancies, in which a maximum is reached at some $v_{o}$ density and then a demagnetization-like process is observed. In this final DFT section, we analyze the magnetic behavior of STF for $x=0.25$ and $\delta=0.125$ i.e., one vacancy/supercell. As we will show in what follows, this is enough to give a step forward in understanding the experimental behavior  and to set the bases for more complex or computationally demanding descriptions.\\

Figure \ref{figure:models} shows the five different vacancies that are found to be symmetrically irreducible for all the Fe-Fe cations configurations, according to our considerations in Section 2. In Table \ref{table3} we summarized all the results for ${{nn_{\mathrm{Fe}}}_{i}}/a \sim 1$, in which FM and AFM states have been relaxed for the corresponding valences of a neutral formula. The sub-indices a$_{1}$ or a$_{2}$ label the vacancies according to Figure \ref{figure:models}. We can see that both a$_{1}$ and a$_{2}$ $gs$ are given by AFM states with what seem to be high Fe spins. Those two systems, though they have the vacancies coordinating a Fe ion, both present different resulting structures i.e., a tetragonal and an orthorhombic-like, respectively, which means that the Fe-$v_{o}$-Fe configuration reduces less the symmetry of the system, what prompts it to be the global ground state among competing a$_{1}^{gs}$ and a$_{2}^{gs}$ as Table \ref{table3} shows. This is also reflected in the magnetic moments, where we see in a$_{2}^{gs}$ two slightly unequal moments differing by $\sim 0.1\mu_{B}$, which is a consequence of the different hybridized $3d$-Fe occupations displayed in Figure \ref{figure: total stoi a1 and a2 and t2g2g a1 and a2}. This last figure shows the total and t$_{2g}$e$_{g}$ densities of states for the $gs$ in the stoichiometric and a$_{1,2}$ deficient cases. For $\delta=0$, the system is FM and both spins are in similar states with polarized seemingly half-filled t$_{2g}$e$_{g}$, which would correspond to a t$_{2g}^{3}$e$_{g}^{1}$ occupancy for Fe$_{h}^{4+}$ spin states. In a$_{1}^{gs}$, Fe cations are AFM ordered and in similar states, i.e., an occupancy close to a t$_{2g}^{4}$e$_{g}^{2}$ for Fe$_{h}^{2+}$ spin states. When checking a$_{2}^{gs}$ in Figure \ref{figure: total stoi a1 and a2 and t2g2g a1 and a2}, we found the solution associated to the mixture of valences spin states, for an AFM ordering, which agrees with the reported ferrimagnetic-like behavior found in STF \cite{OdefSTF}: the Fe ion that lies on the incomplete O$_{5}$ octahedra, here called Fe$_{2}$, seems to have an electronic distribution of a hybridized t$_{2g}^{3}$e$_{g}^{2}$ for a Fe$_{h}^{3+}$ character and therefore larger local magnetic moment than Fe$_{1}$, which resembles the Fe state in the previous cases.\\

One deficient solution for V$_{a_{1}}$ provides us with local annihilation of the magnetization, while a second one V$_{a_{2}}$, provides us with a small magnetization. One can already intuit that one way of increasing or decreasing the perovskite magnetization is to let the system stabilize not just one type of $v_{o}$, which is less plausible experimentally either way, but a group of solutions within a reasonable energy formation. These two vacancies V$_{a_{1,2}}$ happen to have spin-gaps, for FM states to/from AFM ones, just below a few dozen meV, which one could imagine competing with a room temperature excitation threshold for thermal activation up to some extent.\\

%+++++++++++++++++++++++++++++++++++++++++++++++
\begin{figure*}[t]
\center
\vspace{0.0in}
\includegraphics[width=0.99\linewidth]{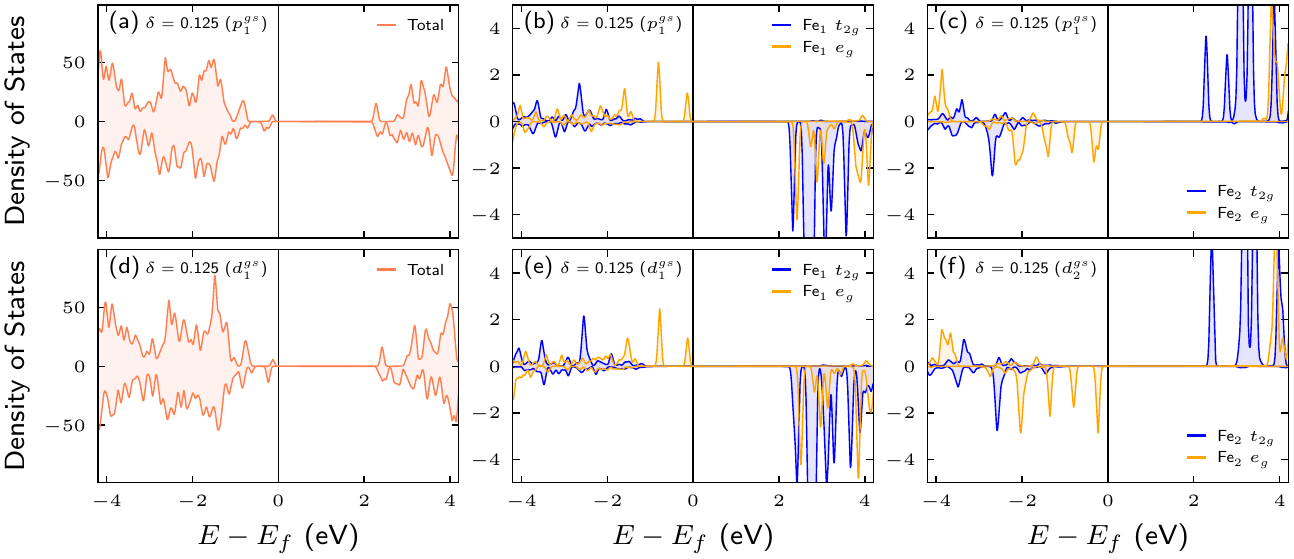}
\vspace{0.0in}
\caption{(a) Total DOS and (b, c) t$_{2g}$e$_{g}$-projected DOS of Fe ions for the V$_{p_{1}}$ vacancy ground state in SrTi$_{0.75}$Fe$_{0.25}$O$_{2.875}$. (d) Total DOS and (e, f) t$_{2g}$e$_{g}$-projected DOS of Fe ions for the V$_{d_{1}}$ vacancy ground state in SrTi$_{0.75}$Fe$_{0.25}$O$_{2.875}$.
\label{figure: total p1 and p2 and t2g2g p1 and p2}}
\end{figure*}
%+++++++++++++++++++++++++++++++++++++++++++++++

We now describe briefly the rest of the $v_{o}$ considered here. Tables \ref{table4} and \ref{table5} present the characterization of the $[1,1,0]$ (plane) and $[1,1,1]$ (diagonal) arrangements, respectively. The first one displays the results corresponding to V$_{p_{1}}$ and V$_{p_{2}}$ in Figure \ref{figure:models}. In this cations distribution the AFM and FM stabilized solutions are energetically close, as the $gs$ for V$_{p_{1,2}}$, both AFM, are within just a couple of meV off the closest FM state, which also present high spin occupation. In both cases, the vacancy coordinates one Fe ion, and therefore one of the Fe magnetic moments is always larger than the other one, in a similar manner to the behavior of Co observed in STC \cite{STC}. In this last case, though, the difference is due to a low-to-high spin state transition, while in STF it originates from hybridization between high Fe$^{4+}$ and Fe$^{3+}$ states as Figure \ref{figure: total p1 and p2 and t2g2g p1 and p2} suggests. Hence, the system relaxes to a magnetized state, either conformed by a ferrimagnetic ordering with $\sim{0.05}\mu_B$/Fe or by a FM one with $\sim{4.16}\mu_B$/Fe, respectively. The energy difference between these orderings seems independent of the V$_{p}$ type as well as the local Fe magnetic moments.\\

Although here we are interested in the magnetic degrees of freedom and V$_{p_{1}}$ and V$_{p_{2}}$ behave similarly from that viewpoint, among those two $v_{o}$ the first one turns out to be more interesting for multiferroic purposes as the $3d$ hybridization is slightly shifted  just below $E_{f}$ and the system is a semiconductor with a $2.15$ eV band-gap, which is adequate for searching of an effective electric polarization in non-metallic ferroelectrics, as it was explored in STFC \cite{STFC_polar}. In this last work, the authors showed how oxygen migrations and deficiency variations can modulate the band-gap and electric polarization in Fe, Co-substituted STO systems with defects such as the ones studied here.\\

Figure \ref{figure: total p1 and p2 and t2g2g p1 and p2} shows the total and projected density of states for V$_{p_{1}}$. It is clear that the system is AFM and that different $3d$-occupancy determines the magnetism of the solid solution, in spite of the close local spin values. One spin is stabilized in a t$_{2g}^{3}$e$_{g}^{1}$ while the other one in a t$_{2g}^{4}$e$_{g}^{2}$. The subtle but consistent up-down population in one of the two t$_{2g}$ orbitals (Fe$_{2}$) of the ground states in Table \ref{table4}, which is also present in several of the less stable ground states, differenciates the two magnetic states of the Fe ions sufficiently as to have an effective magnetic moment/supercell in what is the aforementioned ferrimagnetic ordering. \\

We now analyze $V_{d_{1}}$ in Figure \ref{figure:models}. Table \ref{table5} shows that in this configuration the system stabilizes an AFM ordering. A FM solution ($d_{1}^{ss_{1}}$) is within a very narrow energy difference, which is a repercussion of the large ${{nn_{\mathrm{Fe}}}_{i}}/a \sim\sqrt{3}$ distance compared to an O-mediated $nn_{\mathrm{Fe}}$ super-exchange. Isolate-like Fe ions are coordinated by completed octahedra in two crystallographic directions but [100], creating a tetragonal-pyramid chemical strain that slightly favors indirect antiferromagnetism, in accordance with a quasi-static $d^{4}$-$d^{6}$ $180^{o}$ bond after Goodenough \cite{gerald dionne book}. The second lowest energy state is also AFM though it is not equivalent to $d_{1}^{gs}$. The slightly compact packaging of ${ss}_{1}$ promoted a shifting of the e$_{g}$ in Figure \ref{figure: total p1 and p2 and t2g2g p1 and p2} which creates available hybridized states that prevent the energy band-gap. The magnetic moments are given by high spin states following the behavior exhibited by V$_{p_{1,2}}$, which stabilized t$_{2g}^{3}$e$_{g}^{1}$ and t$_{2g}^{4}$e$_{g}^{2}$ spin states, with the vacancy surrounding the first one (Fe$_{1}$). Also, $d_{1}^{gs}$ has a band-gap of $2.04$ eV, which is larger than the one for the V$_{p}$ case but not necessarily larger than the one for V$_{a_{1}}$. This indicates that ions-proximity is not a predominant factor in the metallic-like effects due to cluster formation in perovskite solid solutions of this kind when we consider $v_{o}$. For instance, in semi-metallic SCO a gap opens up and increases with $\delta$ while perovskites such as Fe, Co-substituted STO present a maximum at intermediate vacancy density \cite{STFC_polar} for different cations $nn$ distances.

%+++++++++++++++++++++++++++++++++++++++++++++++++++++++++++++++++++++++++
%+++++++++++++++++++++++++++++++++++++++++++++++++++++++++++++++++++++++++
\section{MC Results}
We are interested in implementing a model that could give us further insights to understand the change of magnetization when the number of vacancies increases. In this section, we display the results of the Monte Carlo approach described in Section 2. Such a model is able to capture some characteristics of the deficient perovskite magnetism when its magnetic remanence and saturation magnetization are similar, which can be reached by using different substrates and/or tuning the deposition pressures \cite{OdefSTF,STFexp}, in hard and soft magnetic materials \cite{gerald dionne book}.\\

Among all the $v_{o}$ results in Tables \ref{table3}, \ref{table4} and \ref{table5}, we selected a group of vacancies that represent an STF solution with the same magnetic cations arrangement. The supercells in Figure \ref{figure:models} give rise to STF crystals with just one of three Fe-Fe arrangements each; therefore, if the ab-initio energies are approached as statistical weights, those would meaningfully represent a monocrystal i.e., there is no mixture of $[1,0,0]$, $[1,1,0]$ and/or $[1,1,1]$ arranged Fe-Fe magnetic pairs in our scope. Now, to check which vacancies are energetically favorable in each configuration, we calculate the formation energy $E_{f}^{\delta}$ for the $gs$ and $ss$ of each V$_{a,p,d}$. In Table \ref{tablelast2} we can see the formation energy $E_{f}^{\delta}$ of oxygen vacancies as calculated with Equation \ref{formation}:
\begin{equation}
\begin{array}{ccccc}   
 & & E_{f}^{\delta} & = & E_{\delta} - E_{\delta = 0} + \mu_{\mathrm{O}}
\label{formation}
\end{array}
\end{equation}
where $E_{\delta}$ is the total energy of the crystal with one oxygen vacancy after relaxation, and $E_{\delta = 0}$ is the total energy of the stoichiometric crystal ($\delta = 0$) with the corresponding Fe-Fe pair orientation. $\mu_{\mathrm{O}}$ is the chemical potential of O, calculated as half the energy of an isolated  $\mathrm{O}_2$ molecule \cite{SCO_OV}.\\

%+++++++++++++++++
\begin{table}[h]
\caption{$v_{o}$ formation energies $E_f^{\delta}$ and space group (SG) symmetry for gs/ss in Tables \ref{table3}, \ref{table4} and \ref{table5} (HSE06).}
\addtolength{\tabcolsep}{-1pt}
\small
\begin{center}
\begin{tabular}{c c c c c c}
\multicolumn{1}{c}{$E_f^{\delta}$(eV)}  &  V$_{a_1}$ & V$_{a_2}$ &  V$_{p_1}$ & V$_{p_2}$	& V$_{d_1}$\\ 
\hline 
\multicolumn{1}{c}{$gs$} & 3.74 & 4.21  & 3.94 & 3.90 & 3.92 \\
\multicolumn{1}{c}{$ss$} & 3.98 & 4.40 & 3.95 & 3.91 & 3.92 \\
\scriptsize{SG} & \scriptsize{P4/mmm} & \scriptsize{Pmm2} & \scriptsize{P4mm} & \scriptsize{Pmm2} & \scriptsize{P4mm}\\
\hline
\end{tabular}
\end{center}
\label{tablelast2}
\end{table}
%+++++++++++++++++
In the case of $V_{d_{1}}$ we have used ${ss}_{2}$, as ${ss}_{1}$ and $gs$ converged to the same magnetically, and because ${ss}_{2}$ is FM as opposed to $gs$, which happens with the $gs$ and $ss$ for V$_{a,p}$ too. Table \ref{tablelast2} confirms that the diagonal configuration places the Fe ions too far for them to sense the switching of the magnetic coupling and/or slight tetragonal distortions; therefore, we have practically energy-equivalent vacancies for this configuration. Creating these last $v_{o}$ is cheaper than creating V$_{p_{1}}$ but slightly more expensive than V$_{p_{2}}$, which is a consequence of the in-plane structural changes due to the $[1,1,0]$ Fe-ions arrangement that are evenly competing with the Ti-$v_{o}$-Fe chemical strain caused by V$_{p_{2}}$ \cite{STC}, as symmetry in Table \ref{tablelast2} evidences. Among all $v_{o}$ in Table \ref{tablelast2} V$_{a_{1}}$ is energetically cheaper to create although V$_{a_{2}}$ is comparably more expensive. Nevertheless, the $[1,0,0]$ Fe-Fe $\delta=0$ configurations are cheaper overall when revisiting Table \ref{table2}. Moreover, V$_{a_{1}}$ and V$_{a_{2}}$ can be connected by adiabatic oxygen migration paths that support ferroic orders \cite{STFC_polar}, besides that there is a difference between the in-plane and the out-of-plane experimental magnetization in STF, which is related to the influence of magnetoelastic effects that could promote the formation of vertical nanopillars within the perovskite matrix, orienting the resulting magnetization \cite{OdefSTF}. These previous observations leave us with the choice of the ``above" configuration as well as with the energy results in Table \ref{table3} as a reasonable representative STF configurational space to be sampled by our Monte Carlo modeling. \\

%+++++++++++++++++++++++++++++++++++++++++++++++
\begin{figure}[t]
\center
\includegraphics[width=1.0\linewidth]{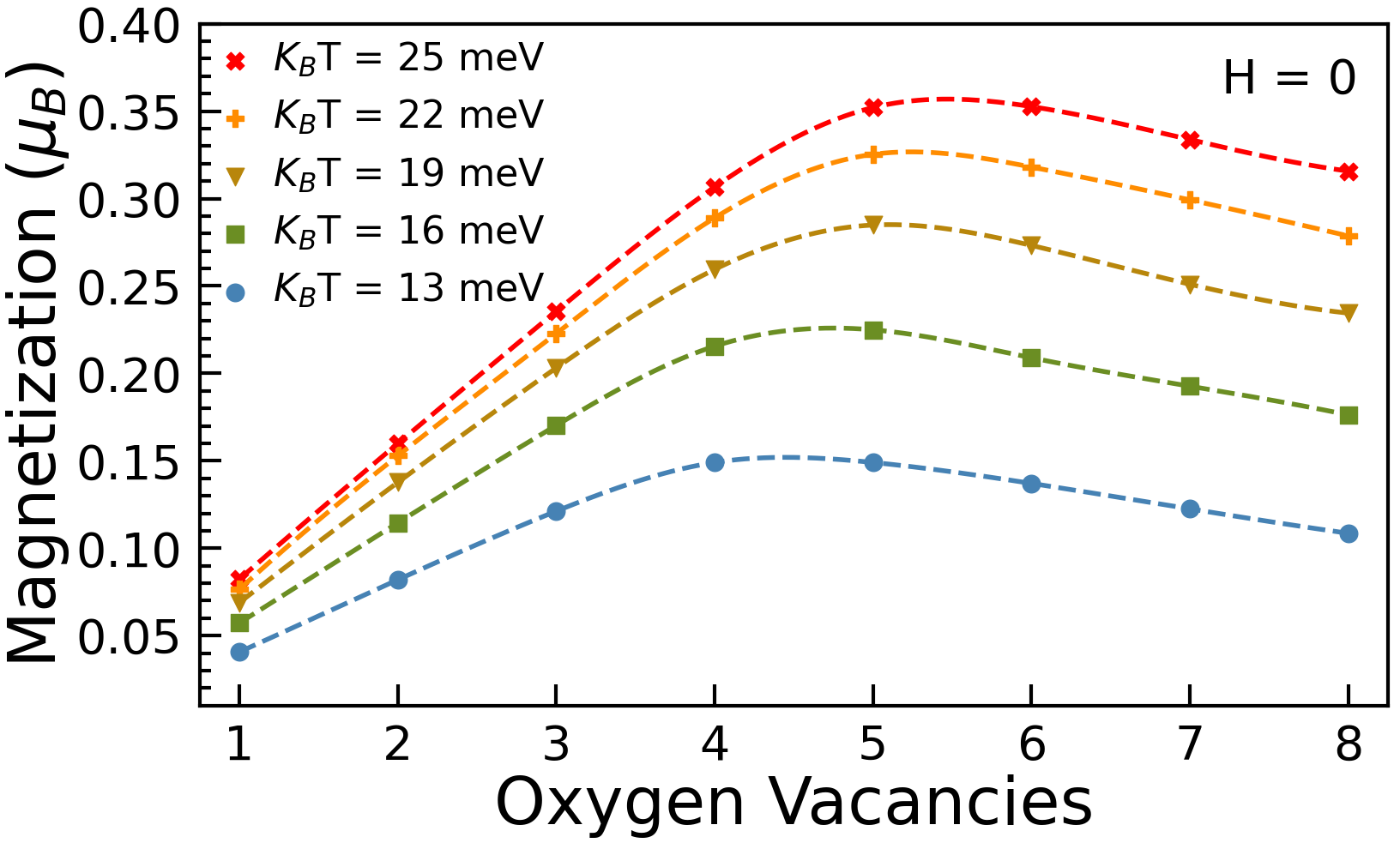}
\caption{DFT based Monte Carlo modeling of oxygen deficient STF ($x=0.25$) magnetization for different number of $v_{o}$. Figure \ref{fig:montecarloalgo} displays the MC scheme implemented to get these magnetizations.}
\label{fig:result2}
\end{figure}
%+++++++++++++++++++++++++++++++++++++++++++++++

We then created vacancies in a $[1,0,0]$ Fe-Fe oriented MC model-crystal starting from a stoichiometric solution. While increasing the number of vacancies such that $\delta$ ranges from $0.0$ to $0.125$, this last value being among the intermediate values at which the ferroic order parameters have been suggested to be maximized for SrTi$_{1-x}$Fe$_{x}$O$_{3}$ with $x=0.25$ \cite{STC,STFC_Ox_hybrid,STFC_polar}, we apply two constrains: (i) $v_{o}$ can take either a V$_{a_{1}}$ or a V$_{a_{2}}$ character/location by bearing the $gs$/$ss$ corresponding energies, associated Fe magnetic moments and corresponding Fe-Fe alignments predicted by our DFT models; (ii) interactions between ``above" Fe-Fe aligned pairs of magnetic cations, separated at least by $2a'$, are negligible. Vacancy interactions in this AB$O_{3-\delta}$ solid solutions as well as charging effects due to these defects are also likely to be negligible as suggested by this and other ab-initio modelling \cite{STFC_Ox_hybrid}.
Finally, we run Monte Carlo simulations by ``throwing" vacancies to the O-octahedral sublattices while following the algorithm illustrated in Figure \ref{fig:montecarloalgo} and the description in Section 2. From the stochastic method, frequencies of specific vacancies-configuration occurrence are obtained, which are then translated to normalized probabilities to be used in Equation \ref{equa:proba}. \\

%+++++++++++++++++++++++++++++++++++++++++++++++
\begin{figure}[t]
\centering
\includegraphics[width=0.9\linewidth]{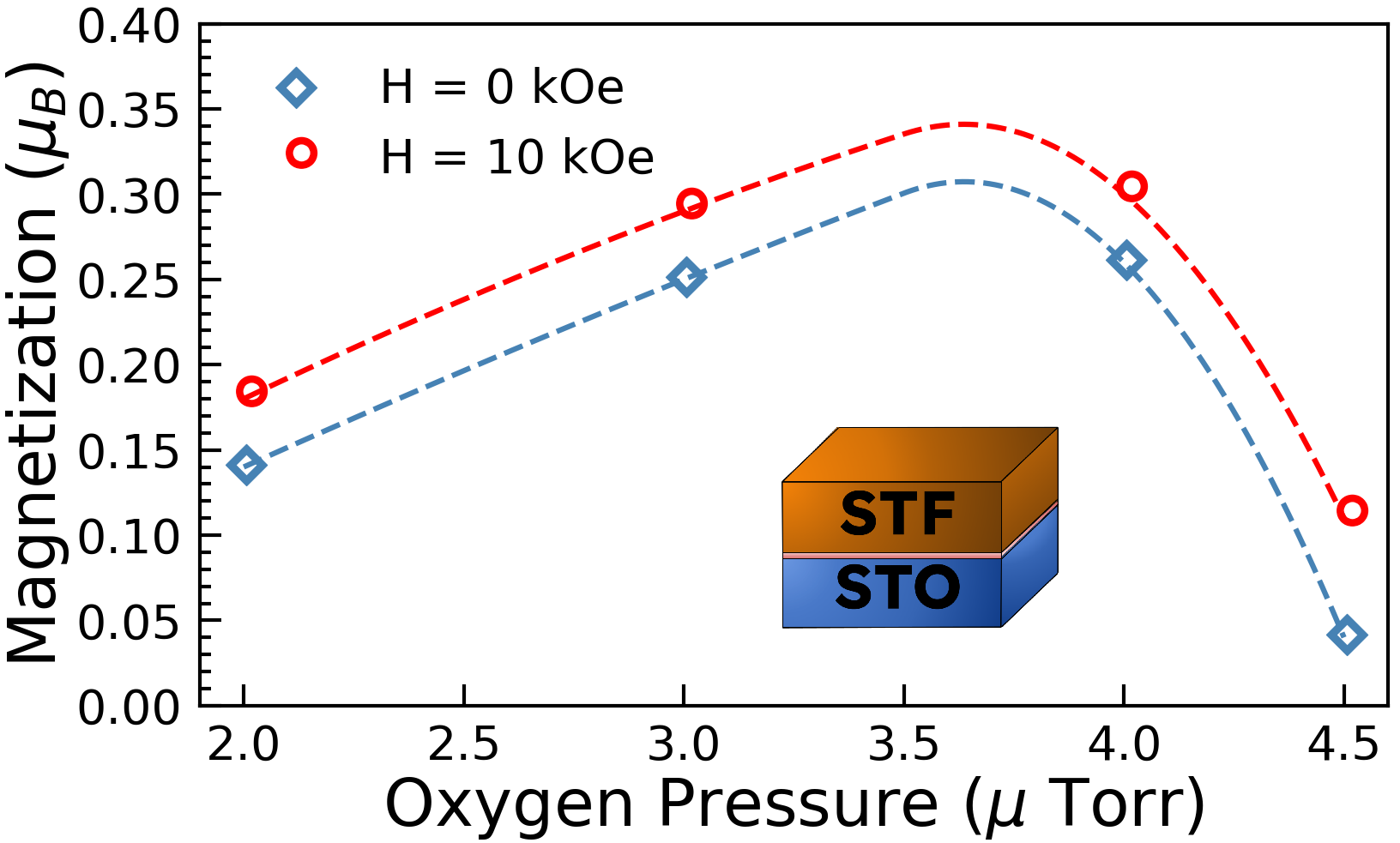}
\includegraphics[width=0.9\linewidth]{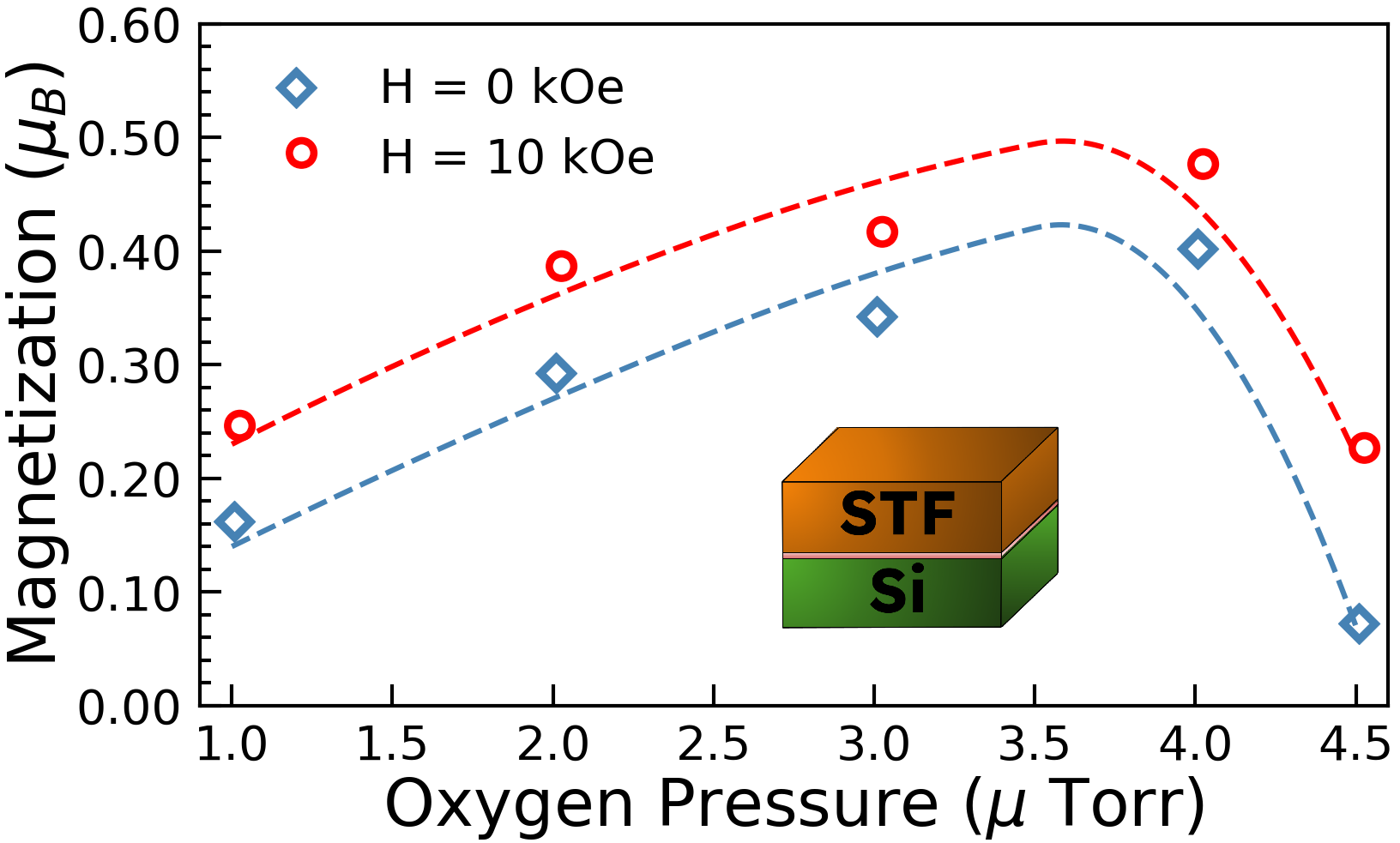}
\caption{Experimental out-of-plane magnetization for oxygen deficient STF films deposited on (a) STO; (b) Si. Selected data reproduced from Ross et al. \cite{OdefSTF} for remanence and saturation magnetizations. Dashed lines are interpolating guiding lines.}
\label{fig:result3}
\end{figure}
%+++++++++++++++++++++++++++++++++++++++++++++++

Figure \ref{fig:result2} displays the modeled magnetization results in the absence of a magnetic field for oxygen deficient STF. For all the adiabatic bath-temperatures simulated here the magnetization has a similar behavior i.e., it increases linearly from low values $\sim{0.05-0.1}\mu_{B}$, to then smoothly reaching a maximum of $\sim{0.35}\mu_{B}$ at intermediate oxygen deficiency $\sim{\delta=0.093}$ ($\sim$ 6 $v_{o}$); then, if the number-of-vacancies/deficiency keeps increasing, such magnetization slowly decreases but with a lesser linear slope than the one it followed to reach its maximum. To compare our intuitive model with experiments, we have extracted from the STF hysteresis loops at room temperature reported in Ross et al. \cite{OdefSTF} the data representing the remanence (H=0 kOe) and saturation magnetization (H=10 kOe). Figure \ref{fig:result3} displays such magnetizations taken at different values of oxygen pressure during deposition. \\

Figure \ref{fig:result3} shows STF magnetization features already remarkably predicted by Figure \ref{fig:result2}. Aforementioned behavior of the magnetization in the model figure is also present in the experimental results. In fact, although the relation between the number/density of vacancies and the oxygen pressure during deposition in these kind of experiments is still an open question and several models have been proposed \cite{STFC_polar,Tuller}, what is clear is that high oxygen pressures mean a lower content of oxygen vacancies and low pressures give rise to large content of $v_{o}$. Therefore, Figures \ref{fig:result2} and Figure \ref{fig:result3} should be read such that the linear behavior below $\sim{5-6}$ $v_{o}$ in the modeling results, which have a faster slope, correspond to the linear behavior for high pressure in the experimental results, which also is the linear part with the faster slope. Slowly decreasing of the magnetization in the modeling when there are more vacancies also represent well the slower decreasing of the magnetization in the experiments, and the threshold given by the saturation magnetization at room temperature ($\sim{25}$ meV) is also captured. \\

If a logarithmic model is used to read the oxygen deficiency $\delta$ out of the oxygen pressure \cite{Tuller}, and it is applied to relate Figure \ref{fig:result2} and Figures \ref{fig:result3}, our model would differ from the experimental results in terms of the location of the maximum of the magnetization by an oxygen deficiency that would be equivalent to a couple of $v_{o}$ i.e., in Figures \ref{fig:result3} the decreasing of the magnetization happens at higher oxygen pressures. One possible reason for this is that the experimental results are performed for a slightly higher Fe content, 0.8 Fe ions/supercell more for $x=0.35$ compared to our stoichiometry, which for an oxygen content given by $3 - x/2 + \delta'$ in Tuller's model \cite{Tuller} means less deficiency $\delta$ for $x=0.35$ in our case, and therefore higher oxygen pressures. Aforementioned model for the oxygen content in STF has been applied at higher temperatures too, nevertheless, Figures \ref{fig:result3} were obtained at room temperature, so to read the relation between oxygen pressure and number of $v_{o}$ with such model-curves would inherently include a temperature shifting effect, which is originated on the increasing of oxygen content for a given pressure when the temperature decreases. In Figure \ref{fig:result2} we showed that the location of the representative maximum of magnetization can be shifted when the temperature is changed. For a given number of vacancies the magnetization decreases if the temperature decreases.\\

%+++++++++++++++++++++++++++++++++++++++++++++++
\begin{figure}[ht]
\centering
\includegraphics[width=0.99\linewidth]{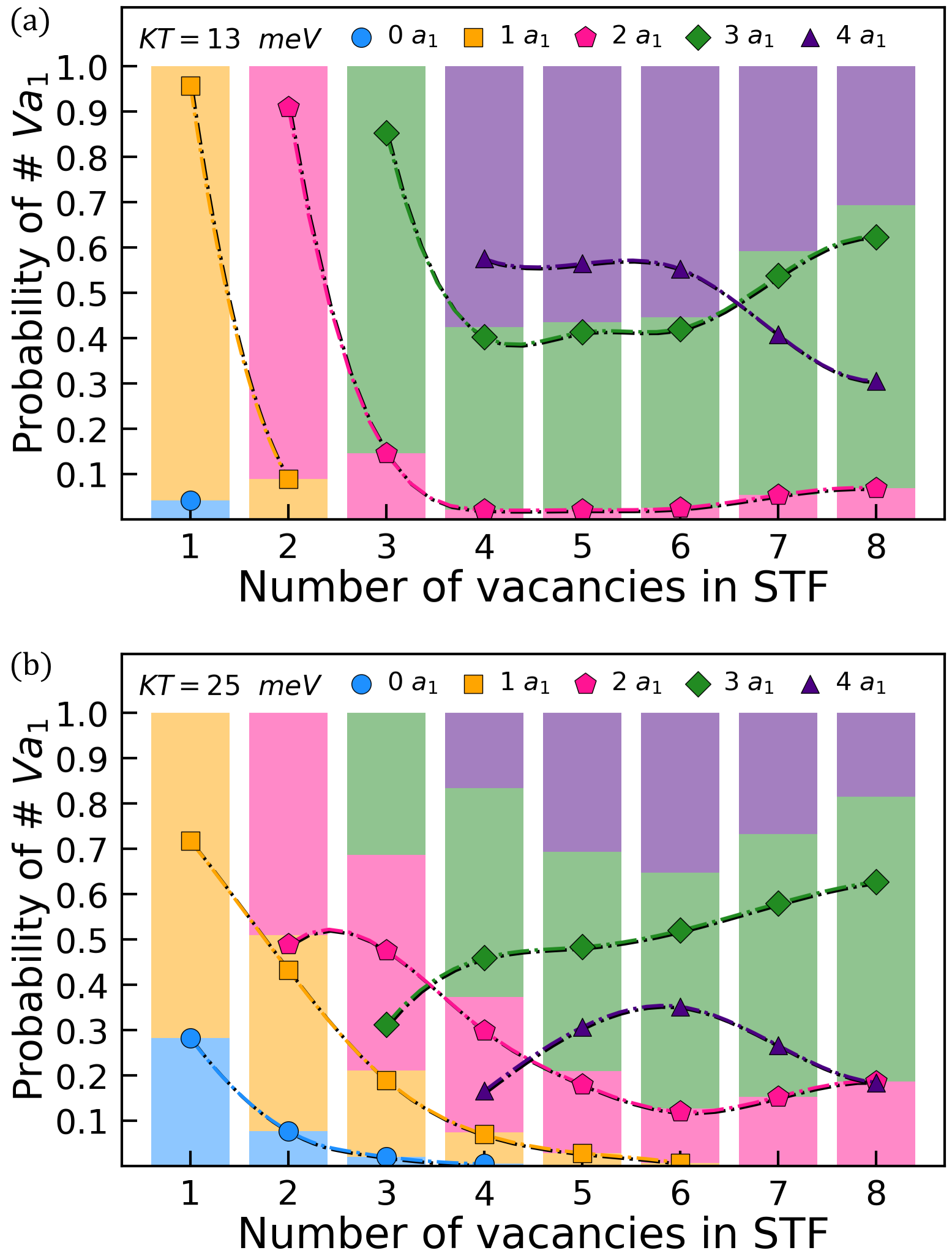}
\caption{Probability of occurrence of specific types of vacancies among V$_{a_{1}}$ and V$_{a_{2}}$ for a given $\delta$ and a bath temperature corresponding to $KT$ e.g., 13 meV (a), and 25 meV (b). For the sake of simplicity, as $P(l)_{V_{a_{1}}} + P(n)_{V_{a_{2}}} = 1$, it is normalized for $(l+n)=\#$ of $v_{o}$ $\Longrightarrow\delta$, we plot just ${P(l)}_{V_{a_{1}}}$ such that the probability ${P(l)}_{V_{a_{1}}}\longrightarrow$ (symbol) $l_{a_{1}}$. We have alternatively used in the background $\%$-bars of the same markers' colors to illustrate the quantity of $l_{a_{1}}$ that would results out of the total number trials in the MC simulations.}
\label{fig:result4}
\end{figure}
%+++++++++++++++++++++++++++++++++++++++++++++++

The ability of our MC simulations to reproduce these previous characteristics is based on the fact that a great portion of the effects observed in Figures \ref{fig:result3} is due to bulk-like magnetic orderings that are thermally activated. In Figure \ref{fig:result4} we have displayed the distribution of the probabilities for the occurrence of a specific number of V$_{a_{1,2}}$ when the temperature is modified. These last figures show that while for low temperatures is less likely to have V$_{a_{2}}$, for room-temperature such probability has increased and therefore the one of having a nonzero magnetization as the V$_{a_{2}}$ has finite magnetization for either FM and AFM polarization; FM ordering increases faster the magnetization though, however, it comes with higher energetic price. When the number of vacancies increases the probability of having less V$_{a_{1}}$ and more V$_{a_{2}}$ increases for a given temperature, this is also because of a topological factor that is rather difficult to measure microscopically i.e., there are more Ti-V$_{a_{2}}$-Fe possible sites than Fe-V$_{a_{1}}$-Fe ones. So, as the bars in Figure \ref{fig:result4} suggests, when there are more possible vacancy sites than actual vacancies and the temperature is low the magnetization increases until a quasi-even distribution between ${a_{1,2}}$ $gs$. Once the number of vacancies competes with the available sites the contribution from ${a_{1}}$ does not increases anymore while ${a_{2}}$ does with mainly antiferromagnetic contributions that are now decreasing the magnetization. This last of course happens slower than the magnetization process with few vacancies that was dominated by FM contributions.  If the temperature increases, it promotes the apparition of new V$_{a_{2}}$ rather than $gs$ to $ss$ transitions in V$_{a}$ as lower panel in Figure \ref{fig:result4} shows, and the magnetization for a sufficient number of vacancies will slowly start to decrease more due to a AFM versus FM competence within V$_{a_{2}}$-given orderings than to the competence between the different types of $v_{o}$.\\

\begin{figure}[t]
\center
\vspace{0.0in}
\includegraphics[width=1\linewidth]{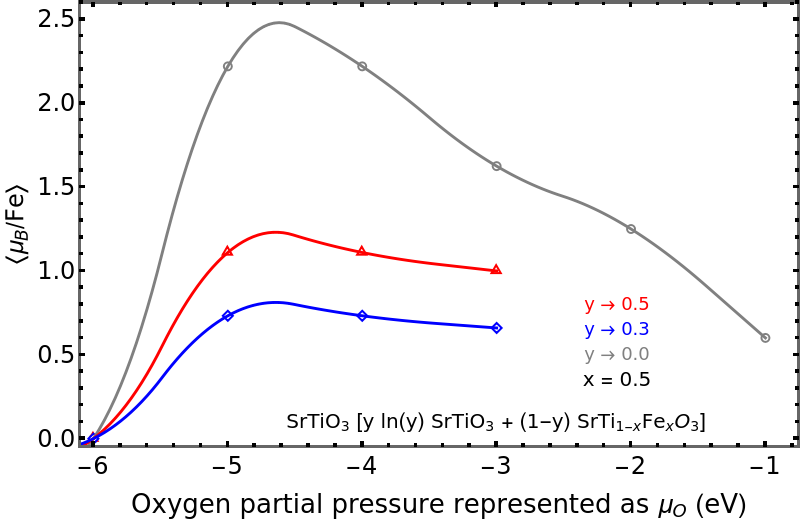}
\vspace{0.0in}
\caption{Magnetism at STO/STF interface versus the oxygen-gas chemical potential $\mu_{O}$. Simulations of the averaged Fe magnetic moment at the resulting interface solution obtained by applying the Interface Reaction of the Materials Project \cite{matproj,materialsproject2,materialsproject3}. Solid colored lines are guiding lines.}
\label{figure:interfacemag}
\end{figure}

Finally, the beautiful balanced competence between what we could call ``Fe-Fe magnetic defects" that we just described is a bulk-like magnetism that dominates the magnetic ordering in these systems. Nonetheless, there is another factor that contributes nicely to the ``quenching" shaping of the magnetization that we observed in Ross's et al. results \cite{OdefSTF} i.e., the magnetism that emerges at the interface between the bulk perovskite and the substrate. Although a deeper study of this part of the system is beyond our scope here, we have asked ourselves what would be the behavior of the magnetic moments at that interface; if in fact it is small compared to the bulk-film, its contribution to the changes of the magnetization' samples would come mainly from the changes that experience the local magnetic moments themselves when the oxygen content is changed and therefore different oxides, among stabilized low/high members of the STF family, contribute with different Fe-stabilized magnetic moments. In Figure \ref{figure:interfacemag} we display the results for the averaged magnetic moments at the interface obtained as described in our Section 2, by applying the Materials Project' Interface Reaction \cite{matproj,materialsproject2,materialsproject3}. In this case we simulate STF/STO interface for three different initial interface compositions and $50\%$ Fe substitution value. This last value was the substitution with more data available at the Materials Project' database for this interface. Figure \ref{figure:interfacemag} shows that the interface material is able to stabilize low spin states, with intermediate values for the oxygen-gas chemical potential being able to mix what could be $3+$ and $2+$ Fe low spins for lower STF content. By increasing Fe content through STF, the system begins to recover the bulk-film behavior, with $4+$ low occupancy being favored. Therefore, the interface also contributes to the ``quenching" shaping of the magnetization by tuning the Fe moments among low spin states, which range within $(0-2) \mu_{B}$. In fact, it is possible to find for large chemical potentials interface-compounds with magnetic contributions arising from Ti ions, which would be in accordance with recent studies in oxygen deficient STO and STFC \cite{STFC_polar} that found magnetic Ti sub-lattices in the perovskite matrices. 

%+++++++++++++++++++++++++++++++++++++++++++++++
%+++++++++++++++++++++++++++++++++++++++++++++++

%+++++++++++++++++++++++++++++++++++++++++++++++++++++++++++++++++++++++++
%+++++++++++++++++++++++++++++++++++++++++++++++++++++++++++++++++++++++++
\section{Conclusions}
We have performed Hybrid-DFT calculations, Monte Carlo simulations as well as reaction-data' magnetic interpretation to model the saturation/remanence magnetization in oxygen deficient STF. We have shown that the experimental behavior is likely due to three main factors: (i) competition between specific ``magnetic defects" conformed by Fe cations that are coordinated by energetically different $v_{o}$, which gives rise to different local spins and magnetic alignments. (ii) The Fe cations distribution in relation to the density of $v_{o}$ plays an important role in defining the oxygen pressure at which the magnetization increasing-process triggered by the AFM to FM local switching starts to be reverted by thermally activated topologically-different ``magnetic defects", which end up giving to the material a ferrimagnetic character. (iii) The magnetic cations at the interface can present a variety of low spin states, in contrast to the high-spin dominated occupancy in the bulk-film material, which on average decreases the magnetic moment per Fe for larger and very low oxygen partial pressures. \\

A full magnetic simulation of this experimental situation would require a many-body Hamiltonian + ab-initio/Molecular dynamics modeling at finite temperature, in which overwhelming simulations for different number of ions would require the microscopic parameters to be calculated self-consistently. It would also require very large supercells if we were to consider all the different configurations for the magnetic ions and vacancies in the structural relaxations at once, which in turn would multiply the configurational space several times as well as the negative magnetic solutions. That would be computationally prohibitive even for some non hybrid functionals. Faster semi-local/meta functionals would require us to constrain the systems to some stabilized valence spin states \cite{allenOccupationMatrixControl2014}, while lacking the compromise between magnetic and conductive properties of hybrids \cite{STFC_Ox_hybrid}. Our approach on the other hand, is able to capture useful characteristics that can be used as fingerprints to synthesize new materials and train Machine Learning based simulations at an affordable computational cost. \\

For a magnet its value of saturation is determinant in terms of e.g., knowing how much magnetic field we need to switch it, as well as its remanence is a witness of the hysteresis process. Our model does not capture hysteresis features, however, we are mostly looking for square-like type of magnets that could be useful for storage/processing information, permanent magnets, energetically efficient switchable systems or with multiferroic features. The model presented here is robust with respect to the representation of the polarized states of a material degrees of freedom that give rise to most characteristics of those aforementioned types of magnets. Moreover, this investigation provides us with hints about the type of crystal symmetry that should be targeted. Thinking in a specific saturation magnetization to be reached for a particular deficiency or working temperature, or vice versa, we could improve on using $\delta$ so we can tailor the magnets saturation features by making use of the properties described here. \\

%+++++++++++++++++++++++++++++++++++++++++++++++++++++++++++++++++++++++++
%+++++++++++++++++++++++++++++++++++++++++++++++++++++++++++++++++++++++++
\section{Acknowledgements}
J.M. Florez and E.S. Morell thank support from project FONDECYT Regular 1221301, Agencia Nacional de Investigación y Desarrollo (ANID), Chile. M.A. Solis thanks financial support of USM-DPP, Chile. C.A. Ross thanks National Science Foundation: DMR 1419807. \\

J.M. Florez also thanks Prof. Shyue P. Ong from the Department of NanoEngineering at the University of California, San Diego, for helpful discussions about the Materials Project' tools for oxygen deficient perovskites modeling.  

%+++++++++++++++++++++++++++++++++++++++++++++++++++++++++++++++++++++++++
%+++++++++++++++++++++++++++++++++++++++++++++++++++++++++++++++++++++++++
\clearpage
\onecolumn
\section*{Appendices}
\appendix
%+++++++++++++++++++++++++++++++++++++++++++++++
\section{STF for $\delta=0.125$ and $x=0.25$}
\label{apen3}
%+++++++++++++++++
\begin{table*}[h!]
\caption{Sr$^{2+}$Ti$^{4+}_{0.75}$Fe$^{y}_{0.25}$O$^{2-}_{2.875}$ properties for vacancies $\bf V_{a_{1}}$ and $\bf V_{a_{2}}$ (HSE06). FM and AFM energies of $nn_{\mathrm{Fe}}\sim{a'}$ configurations for each initial ${S_{i}^{1,2}}$ Fe spins, as well as lattice parameters, magnetic structure, band-gap and relative energy of the selected $gs$ and $ss$ states.}
\begin{center}
\begin{tabular}{cccccc}
            \hline
			\multirow{3}{*}{$y$} & \multirow{3}{*}{$S_{i}^{1,2}$}  & \multicolumn{2}{c}{ $\bf V_{a_{1}}$ } & \multicolumn{2}{c}{ $\bf V_{a_{2}}$ } 
			\\
            \cmidrule(lr){3-4} \cmidrule(lr){5-6}

			&  & $E_{FM}$ & $E_{AFM}$ & $E_{FM}$ & $E_{AFM}$ \\
			&  & (meV/f.u.) & (meV/f.u.) & (meV/f.u.) & (meV/f.u.) \\\hline

            \multirow{4}{*}{(2+,4+)} 
			& $h,h$ & 248.3 & ---$_{gs}$ & 98.9 &  0.02\\
			& $h,l$ & 222.5 & 107.9  & 223.7 & 134.9\\
			& $l,h$ & 236.1 & 233.7 & 206.6 & 197.6\\
			& $l,l$ & 108.3  & 108.4  & 78.8 & 135.3\\
            \hline
			\multirow{3}{*}{(4+,2+)} 
			& $h,l$ & 238.1 & 232.8 & 197.6 & 197.5\\
			& $l,h$ & 222.8 & 107.9  & 234.8 & 135.6\\
			& $l,l$ & 107.9  & 108.0  & 135.2 & 135.2\\
			\hline
			\multirow{4}{*}{(3+,3+)} 
			& $h,h$ & 31.2$_{ss}$  & 0.06   & 23.8$_{ss}$& ---$_{gs}$\\ 
			& $h,l$ & 313.3 & 234.8 & 268.6 & 197.4\\
			& $l,h$ & 313.3 & 233.3 & 196.9 & 197.6\\
			& $l,l$ & 447.9 & 218.2 & 408.3 & 0.02\\ 
   
            \midrule
			State & $S_{f}^{1,2} (\mu_B)$ & $(a',b',c')$ (\AA) & $V$  (\AA$^3$) & $E_{bg}$ (eV) &  ${\Delta E_{a_1^{gs}}}$ (meV/f.u.)\\ \hline
			
			\bf ${a_{1}}^{gs}$ & (4.07, -4.07) & (7.82, 7.82, 7.83) & 479.33 & 2.15 & ---\\
			\bf ${a_{1}}^{ss}$ & (4.14, 4.14) & (7.82, 7.82, 7.84) & 479.65 & 0 & 31.2\\\hline
			\bf ${a_{2}}^{gs}$ & (4.07, -4.17) & (7.83, 7.85, 7.83) & 481.48 & 0 & 59.6\\
			\bf ${a_{2}}^{ss}$ & (4.14, 4.22) & (7.84, 7.84, 7.84) & 481.72 & 0 & 83.4\\
\hline
\end{tabular}
\end{center}
\label{table3}
\end{table*}
%+++++++++++++++++

%+++++++++++++++++
\begin{table*}[hbt!]
\caption{Sr$^{2+}$Ti$^{4+}_{0.75}$Fe$^{y}_{0.25}$O$^{2-}_{2.875}$ properties for vacancies $\bf V_{p_{1}}$ and $\bf V_{p_{2}}$ (HSE06). FM and AFM energies of $nn_{\mathrm{Fe}}\sim{a'\sqrt{2}}$ configurations for each initial ${S_{i}^{1,2}}$ Fe spins, as well as lattice parameters, magnetic structure, band-gap and relative energy of the selected $gs$ and $ss$ states.}
\begin{center}
\begin{tabular}{cccccc}
        \hline
        \multirow{3}{*}{$y$} & \multirow{3}{*}{$S_{i}^{1,2}$}  & \multicolumn{2}{c}{ $\bf V_{p_{1}}$ } & \multicolumn{2}{c}{ $\bf V_{p_{2}}$ } 
        \\
        \cmidrule(lr){3-4} \cmidrule(lr){5-6}
        &  & $E_{FM}$ & $E_{AFM}$ & $E_{FM}$ & $E_{AFM}$ \\
        &  & (meV/f.u.) & (meV/f.u.) & (meV/f.u.) & (meV/f.u.) \\\hline

        \multirow{4}{*}{(2+,4+)} 
        & $h,h$ & 112.4 &  0.001  & 94.0 & 0.001\\
        & $h,l$ & 166.1 &  128.9 & 275.0 & 143.4\\
        & $l,h$ & 230.9 &  165.6 & 210.7 & 193.2\\
        & $l,l$ & 371.7 &  129.3 & 93.0 & 143.4\\\hline
        \multirow{3}{*}{(4+,2+)}
        & $h,l$ &  165.5 & 165.6 & 193.6 & 193.4\\
        & $l,h$ & 217.4 &  129.2 & 319.8 & 142.0\\
        & $l,l$ & 143.5 & 129.5 & 143.4 & 143.4\\
        \hline
        \multirow{4}{*}{(3+,3+)} 
        & $h,h$ &0.8$_{ss}$  &  0.002 & 0.6$_{ss}$ & 0.008\\ 
        & $h,l$ &  166.1 & 165.6 & 273.7 & 194.0\\
        & $l,h$ & 209.4 & 165.6 & 223.7 & 193.2\\
        & $l,l$ & 371.8 & --- $_{gs}$ & 400.4 &---$_{gs}$\\ 
        \hline 
        		
        State & $S_{f}^{1,2} (\mu_B)$ & $(a',b',c')$ (\AA) & $V$  (\AA$^3$) & $E_{bg}$ (eV) &  ${\Delta E_{p_2^{gs}}}$ (meV)\\ \hline
		
		\bf ${p_{1}}^{gs}$ & (4.12, -4.19) & (7.83, 7.83, 7.86) & 481.38 & 1.94 & 5.1\\
		\bf ${p_{1}}^{us}$ & (4.12, 4.19) & (7.83, 7.83, 7.85) & 481.51 & 0 & 5.9\\\hline
		\bf ${p_{2}}^{gs}$ & (4.12, -4.20) & (7.84, 7.85, 7.83) & 481.57 & 0 &     ---\\
		\bf ${p_{2}}^{us}$ & (4.12, 4.20) & (7.84, 7.84, 7.83) & 481.25 & 0 & 0.6\\
		\hline
\end{tabular}
\end{center}
\label{table4}
\end{table*}
%+++++++++++++++++

%+++++++++++++++++
\begin{table*}[hbt!]
\caption{Sr$^{2+}$Ti$^{4+}_{0.75}$Fe$^{y}_{0.25}$O$^{2-}_{2.875}$ properties for vacancy $\bf V_{d_{1}}$ (HSE06). FM and AFM energies of $nn_{\mathrm{Fe}}\sim{a'\sqrt{3}}$ configurations for each initial ${S_{i}^{1,2}}$ Fe spins, as well as lattice parameters, magnetic structure, band-gap and relative energy of the selected $gs$ and $ss$ states.}
\begin{center}
\begin{tabular}{cccccc}
		\hline
        \multirow{2}{*}{$y$} & \multirow{2}{*}{$S_{i}^{1,2}$} & \multicolumn{2}{c}{$E_{FM}$}  & \multicolumn{2}{c}{$E_{AFM}$} \\
        &  & \multicolumn{2}{c}{(meV/f.u.)}  & \multicolumn{2}{c}{(meV/f.u.)} \\\hline

  		\multirow{4}{*}{(2+,4+)} 
		& $h,h$ & \multicolumn{2}{c}{105.1} &  \multicolumn{2}{c}{--- $_{gs}$}  \\
		& $h,l$ & \multicolumn{2}{c}{306.9} &  \multicolumn{2}{c}{132.4} \\
		& $l,h$ & \multicolumn{2}{c}{215.5} & \multicolumn{2}{c}{172.5}  \\
		& $l,l$ & \multicolumn{2}{c}{102.9} &  \multicolumn{2}{c}{132.7} \\\hline
		\multirow{3}{*}{(4+,2+)}
		& $h,l$ & \multicolumn{2}{c}{273.3} & \multicolumn{2}{c}{172.3} \\
		& $l,h$ & \multicolumn{2}{c}{231.0} & \multicolumn{2}{c}{132.9} \\
		& $l,l$ & \multicolumn{2}{c}{144.0} & \multicolumn{2}{c}{132.4} \\
		\hline
		\multirow{4}{*}{(3+,3+)} 
		& $h,h$ & \multicolumn{2}{c}{0.04$_{ss_2}$}  & \multicolumn{2}{c}{0.1} \\ 
		& $h,l$ & \multicolumn{2}{c}{174.2} & \multicolumn{2}{c}{172.3} \\
		& $l,h$ & \multicolumn{2}{c}{215.2} & \multicolumn{2}{c}{172.3} \\
		& $l,l$ & \multicolumn{2}{c}{382.9} & \multicolumn{2}{c}{0.1$_{ss_1}$} \\\hline

		State & $S_{f}^{1,2} (\mu_B)$ & \multicolumn{2}{c}{$(a',b',c')$ (\AA)} & $V$ (\AA$^3$) & $E_{bg}$ (eV) \\ \hline
        \bf $d_1^{gs}$ & (4.12, -4.12) & \multicolumn{2}{c}{(7.83, 7.83, 7.85)}  & 481.83 & 2.04 \\
		\bf $d_1^{ss_1}$ & (4.12, -4.12) & \multicolumn{2}{c}{(7.83, 7.83, 7.85)}  & 481.51 & 0 \\
		\bf $d_1^{ss_2}$ & (4.12, 4.12) & \multicolumn{2}{c}{(7.83, 7.83, 7.85)}  & 481.62 & 0\\
        \hline
\end{tabular}
\end{center}
\label{table5}
\end{table*}
%+++++++++++++++++

%+++++++++++++++++++++++++++++++++++++++++++++++
\clearpage
\section{Monte Carlo modeling}
\label{apen4}

%+++++++++++++++++
\begin{figure}[ht]
\centering
\includegraphics[width=1.0\linewidth]{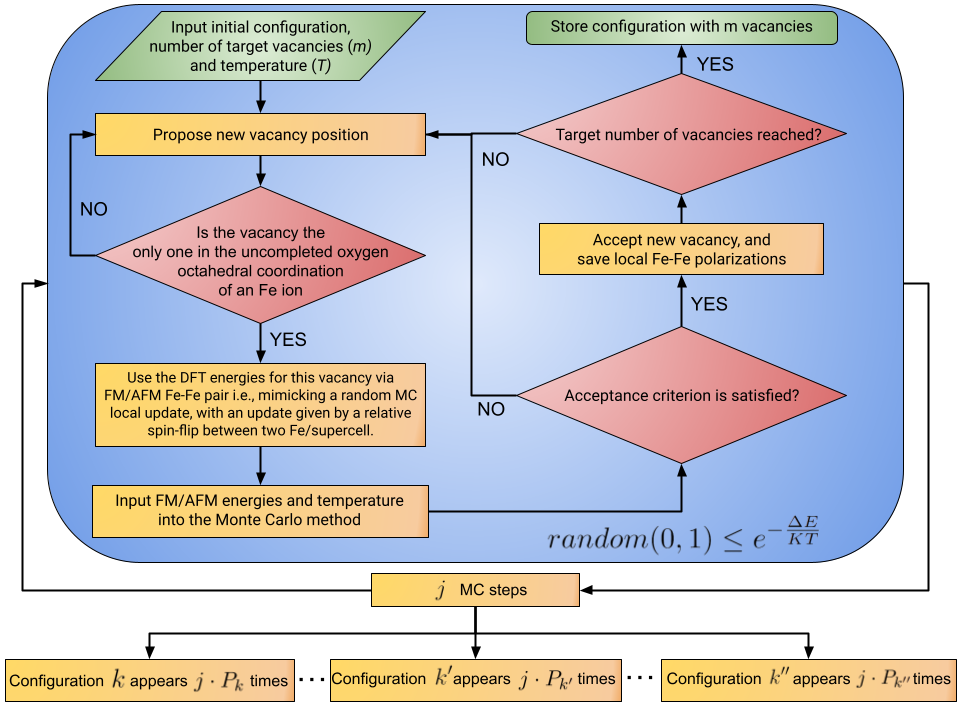}
\caption{Diagram of the algorithm used to implement a Metropolis Monte Carlo scheme, which generates effective magnetic sites out of pairs of Fe ions constrained by uncompleted coordinating oxygen-octahedra. Such effective magnetic moments are averaged for every possible vacancy distribution to then calculate the spontaneous perovskite magnetization by using the probabilities of occurrence $P_{k}$ of such distributions along with their configurational magnetization $M_{k}$, through Equation \ref{equa:proba}.}
\label{fig:montecarloalgo}
\end{figure}
%+++++++++++++++++

%+++++++++++++++++++++++++++++++++++++++++++++++++++++++++++++++++++++++++++++++++++++
%+++++++++++++++++++++++++++++++++++++++++++++++++++++++++++++++++++++++++++++++++++++
%\bibliography{ref.bib}

\section*{References}
\begin{thebibliography}{100}
\bibitem{Ferrooxide_superla} G. Rijnders and D. H. A. Blank, Nature 433, 369 EP (2005).
%Build your own superlattice. Artificial materials made from oxide building blocks turn out to be excellent ferroelectrics. 

\bibitem{BiFeO_multi_hetero} J. Wang, J. B. Neaton, H. Zheng, V. Nagarajan, S. B. Ogale, B. Liu, D. Viehland, V. Vaithyanathan, D. G. Schlom, U. V. Waghmare, N. A. Spaldin, K. M. Rabe, M. Wuttig, and R. Ramesh, Science 299, 1719 (2003).
%Epitaxial BiFeO3 Multiferroic Thin Film Heterostructures 

\bibitem{BaTiOFerroel_film}  K. J. Choi, M. Biegalski, Y. L. Li, A. Sharan, J. Schubert, R. Uecker, P. Reiche, Y. B. Chen, X. Q. Pan, V. Gopalan, L.-Q. Chen, D. G. Schlom, and C. B. Eom, Science 306, 1005 (2004).
%Enhancement of Ferroelectricity in Strained BaTiO Thin Films

\bibitem{Ferroelec_strin_sto} H. W. Jang, A. Kumar, S. Denev, M. D. Biegalski, P. Maksymovych, C. W. Bark, C. T. Nelson, C. M. Folkman, S. H. Baek, N. Balke, C. M. Brooks, D. A. Tenne, D. G. Schlom, L. Q. Chen, X. Q. Pan, S. V. Kalinin, V. Gopalan, and C. B. Eom, Phys. Rev. Lett. 104, 197601 (2010).
%Ferroelectricity in Strain-Free SrTiO3 Thin Films
\bibitem{superiso_sto} A. Stucky, G. W. Scheerer, Z. Ren, D. Jaccard, J.M. Poumirol, C. Barreteau, E. Giannini, and D. van der Marel, Scientific Reports 6, 37582 EP (2016).
%Isotope effect in superconducting n-doped SrTiO3
%5 papers

\bibitem{supersemi_sto} J. F. Schooley, W. R. Hosler, and M. L. Cohen, Phys. Rev. Lett. 12, 474 (1964).
%Superconductivity in semiconducting SrTiO3

\bibitem{superinter_bet_oxi} N. Reyren, S. Thiel, A. D. Caviglia, L. F. Kourkoutis, G. Hammerl, C. Richter, C. W. Schneider, T. Kopp, A.S. Raetschi, D. Jaccard, M. Gabay, D. A. Muller, J.M. Triscone, and J. Mannhart, Science 317, 1196 (2007).

\bibitem{nonstoi_grain_sto} Miyoung Kim, Gerd Duscher, Nigel D. Browning, Karl Sohlberg, Sokrates T. Pantelides, and Stephen J. Pennycook. Phys. Rev. Lett. 86, 4056 4059 (2001). %Nonstoichiometry and the Electrical Activity of Grain Boundaries in SrTiO3

\bibitem{multiferroico1} M. Fiebig, T. Lottermoser, D. Meier, and M. Trassin, Nature Reviews Materials 1, 16046 (2016).  
%The evolution of multiferroics

\bibitem{OdefSTF} T. Goto, D. H. Kim, X. Sun, M. C. Onbasli, J. M. Florez, S. P. Ong, P. Vargas, K. Ackland, P. Stamenov, N. M. Aimon, M. Inoue, H. L. Tuller, G. F. Dionne, J. M. Coey, and C. A. Ross, Phys. Rev. Applied 7, 024006 (2017).
%Magnetism and Faraday Rotation in Oxygen-Deficient Polycrystalline and Single-Crystal Iron-Substituted Strontium Titanate 

\bibitem{multiferro_roomt_ortho-morpho_lfo}  S. Song, H. Han, H. M. Jang, Y. T. Kim, N.-S. Lee, C. G. Park, J. R. Kim, T. W. Noh, and J. F. Scott, Advanced Materials 28, 7430 (2016).
%Implementing Room-Temperature Multiferroism by Exploiting Hexagonal-Orthorhombic Morphotropic Phase Coexistence in LuFeO3 Thin Films

\bibitem{Ferroindu_isoto_exchan_sto} M. Itoh, R. Wang, Y. Inaguma, T. Yamaguchi, Y.-J. Shan, and T. Nakamura, Phys. Rev. Lett. 82, 3540 (1999).
%Ferroelectricity Induced by Oxygen Isotope Exchange in Strontium Titanate Perovskite

\bibitem{anti-ferrodist_vacan_sto} M. Choi, F. Oba, Y. Kumagai, and I. Tanaka, Advanced Materials 25, 86 (2013).
%Anti-ferrodistortive-Like Oxygen-Octahedron Rotation Induced by the Oxygen Vacancy in Cubic SrTiO3

\bibitem{oxy_vac_stco_teo-exp} C. Mitra, C. Lin, A. B. Posadas, and A. A. Demkov, Phys. Rev. B 90, 125130 (2014).
%Role of oxygen vacancies in room-temperature ferromagnetism in cobalt-substituted SrTiO3

\bibitem{pointdefect_ferroelec_sto} K. Klyukin and V. Alexandrov, Phys. Rev. B 95, 035301 (2017).
%Effect of intrinsic point defects on ferroelectric polarization behavior of SrTiO3

\bibitem{large_magne_Odefici_sto} A. Lopez-Bezanilla, P. Ganesh, and P. B. Littlewood, APL Materials 3, 100701 (2015).
%Research Update: Plentiful magnetic moments in oxygen deficient SrTiO3

\bibitem{electro_doping_metal_oxide} A. Cammarata and J.M. Rondinelli, Applied Physics Letters 108, 213109 (2016).
%Electronic doping of transition metal oxide perovskites

\bibitem{Sikam:2018daa} Brovko, O. O., and Tosatti, E., Physical Review Materials, 1(4), 0444058  (2017). 
%Controlling the magnetism of oxygen surface vacancies in SrTiO3 through charging.


\bibitem{sto_with_hse} F. El-Mellouhi, E. N. Brothers, M. J. Lucero, and G. E. Scuseria, Phys. Rev. B 84, 199907 (2011).
%Erratum: Modeling of the cubic and antiferrodistortive phases of SrTiO3 with screened hybrid density functional theory [Phys. Rev. B 84, 115122 (2011)]

\bibitem{Pai:2018fia} Pai, Y.-Y., Tylan-Tyler, A., Irvin, P., and Levy, J., Reports on Progress in Physics, 81(3), 03650 (2018). 
%Physics of SrTiO 3-based heterostructures and nanostructures: a review.

\bibitem{MagSTOdelta} Y. Zhang, J. Wang, M. Sahoo, T. Shimada, and T. Kitamura, Phys. Chem. Chem. Phys. 17, 27136 (2015).
%Mechanical control of magnetism in oxygen deficient perovskite SrTiO3

\bibitem{OrbitalSymmetrySTO} C. Lin, C. Mitra, and A. A. Demkov, Phys. Rev. B 86, 161102 (2012).
%Orbital ordering under reduced symmetry in transition metal perovskites:Oxygen vacancy in SrTiO3

\bibitem{StrainControl} J. R. Petrie, C. Mitra, H. Jeen, W. S. Choi, T. L. Meyer, F. A. Reboredo, J. W. Freeland, G. Eres, and H. N. Lee, Advanced Functional Materials 26, 1564 (2016).
%Strain Control of Oxygen Vacancies in Epitaxial Strontium Cobaltite Films

\bibitem{Dong:2018dr} Dong, X.-L., Zhang, K.-H., and Xu, M.-X., Frontiers of Physics, 13(5), 1155037 (2018). 
%First-principles study of electronic structure and magnetic properties of SrTi1-xMxO3 (M = Cr, Mn, Fe, Co, or Ni).

\bibitem{Lee:2018dka} Lee, D., Wang, H., Noesges, B. A., Asel, T. J., Pan, J., Lee, J. W., et al., Physical Review Materials, 2(6), 060403 (2018). 
%Identification of a functional point defect in ${\rm{SrTi}}{{\rm{O}}_3}$, 17.

\bibitem{Schiaffino:2017cwa} Sikam, P., Moontragoon, P., Sararat, C., Karaphun, A., Swatsitang, E., Pinitsoontorn, S., and Thongbai, P., Applied Surface Science, 446, 92113 (2018). 
%DFT calculation and experimental study on structural, optical and magnetic properties of Co-doped SrTiO3.

\bibitem{Brovko:2017hh} Wang, Y.-G., Tang, X.-G., Liu, Q.-X., Jiang, Y.-P., and Jiang, L.-L., Nanomaterials, 7(9), 26412 (2017). 
%Room Temperature Tunable Multiferroic Properties in Sol-Gel-Derived Nanocrystalline Sr(Ti1-xFex) Thin Films.

\bibitem{Wang:2017ii} Zhang, Y., Kurt, O., Ascienzo, D., Yang, Q., Le, T., Greenbaum, S., et al., Journal of Physical Chemistry C, 122(24), 12864 12868 (2018).
%Detection of Nanoscale Structural Defects in Degraded Fe-Doped SrTiO3 by Ultrafast Photoacoustic Waves.

\bibitem{gerald dionne book} Dionne, G. F., New York: Springer, (2009).
% Magnetic oxides (Vol. 14, p. 15)

\bibitem{STFexp} D. H. Kim, N. M. Aimon, L. Bi, J. M. Florez, G. F. Dionne, and C. A. Ross, Journal of Physics: Condensed Matter 25, 026002 (2013).
%Magnetostriction in epitaxial SrTi1ÂÂxFexO3 perovskite films with x = 0.13 and 0.35

\bibitem{STC} J. M. Florez, S. P. Ong, M. C. Onbsli, G. F. Dionne, P. Vargas, G. Ceder, and C. A. Ross, Applied Physics Letters 100, 252904 (2012).
%First-principles insights on the magnetism of cubic SrTi12xCoxO32d

\bibitem{oxidation_partialcharge_ionicity} A. Walsh, A. A. Sokol, J. Buckeridge, D. O. Scanlon, and C. R. A. Catlow, The Journal of Physical Chemistry Letters 8, 2074 (2017).
%Electron Counting in Solids: Oxidation States, Partial Charges, and Ionicity

\bibitem{STFC_Ox_hybrid} M. A. Opazo, S. P. Ong, P. Vargas, C. A. Ross, and J. M. Florez, Phys. Rev. Mater., 3, 014404 (2019).
%Oxygen-vacancy tuning of magnetism in SrTi$_{0.75}$Fe$_{0.125}$Co$_{0.125}$O$_{3-\delta}$ perovskite

\bibitem{STFC_polar} Cortés Estay, E. A., Ong, S. P., Ross, C. A.,  Florez, J. M.,  Magnetochemistry, 8(11), 144, (2022).
%Oxygen Deficiency and Migration-Mediated Electric Polarization in Magnetic Fe, Co-Substituted SrTiO3− δ.

\bibitem{STO_Review}  Y. Pai, A. Tylan-Tyler, P. Irvin, and J. Levy, Rep. Prog. Phys. 81, 036503 (2018).
%Physics of SrTiO$_3$-based heterostructures and nanostructures: a review

\bibitem{STO_comp_AFD-FE} U. Aschauer and N. A. Spaldin, J. Phys.: Condens. Matter 26, 122203 (2014).
%Competition and cooperation between antiferrodistortive and ferroelectric instabilities in the model perovskite SrTiO$_3$,

\bibitem{STO_SrOO_vac}  Y. S. Kim, J. Kim, S. J. Moon, W. S. Choi, Y. J. Chang, J.-G. Yoon, J. Yu, J.-S. Chung, and T. W. Noh, Appl. Phys. Lett. 94, 202906 (2009).
%Localized electronic states induced by defects and possible origin of ferroelectricity in strontium titanate thin films,

\bibitem{STO_Ti_antisite} M. Choi, F. Oba, and I. Tanaka, Phys. Rev. Lett. 103, 185502 (2009).
%Role of Ti Antisitelike Defects in SrTiO$_3$,

\bibitem{STO_Pol_defects} K. Klyukin and V. Alexandrov, Phys. Rev. B 95, 035301 (2017).
%Effect of intrinsic point defects on ferroelectric polarization behavior of SrTiO$_3$,

\bibitem{STO_FE_SrTi_ratio}  F. Yang, Q. Zhang, Z. Yang, J. Gu, Y. Liang, W. Li, W. Wang, K. Jin, L. Gu, and J. Guo, Appl. Phys. Lett. 107, 082904 (2015).
% Room-temperature ferroelectricity of SrTiO3 films modulated by cation concentration,

\bibitem{STO111} I. Hallsteinsen, M. Nord, T. Bolstad, P.-E. Vullum, J. E. Boschker, P. Longo, R. Takahashi, R. Holmestad, M. Lippmaa, and T. Tybell, Cryst. Growth Des. 16, 2357 (2016).
%Effect of Polar (111)-Oriented SrTiO$_3$ on Initial Perovskite Growth,

\bibitem{STO001} I. Sokolovic, M. Schmid, U. Diebold, and M. Setvin, Phys. Rev. Mater. 3, 034407 (2019).
%Incipient ferroelectricity: A route towards bulk-terminated,

\bibitem{ferro6} El-Naser, A. A. et al, Philos Mag 101, 1–19 (2020).
% Study the influence of oxygen-deficient (delta = 0.135) in SrFeO3-delta nanoparticles perovskite on structural, electrical and magnetic properties.

\bibitem{STF_multiferroic2} X. Wang, Z. Wang, Q. Hu, C. Zhang, D. Wang, and L. Li, Solid State Commun. 289, 22 (2019).
% Room temperature multiferroic properties of Fe-doped nonstoichiometric SrTiO$_3$ ceramics at both A and B sites,

\bibitem{STC2018} A. S Tang, M. C. Onbasli, X. Sun, and C. A. Ross, ACS Appl. Mater. Interfaces, 10, 7469 (2018).
% Thickness-Dependent Double-Epitaxial Growth in Strained SrTi$_{0.7}$Co$_{0.3}$O$_{3-\delta}$ Films

\bibitem{roomt1} D. H. Kim, L. Bi, P. Jiang, G. F. Dionne, and C. A. Ross, Phys. Rev. B 84, 014416 (2011).
%Magnetoelastic effects in SrTi1xMxO3 (M = Fe, Co, or Cr) epitaxial thin films

\bibitem{roomt2} L. Bi, H.-S. Kim, G. F. Dionne, and C. A. Ross, New Journal of Physics 12, 043044 (2010).
%Structure, magnetic properties and magnetoelastic anisotropy in epitaxial Sr(Ti1xCox)O3 films

\bibitem{FMSTCdelta} A. B. Posadas, C. Mitra, C. Lin, A. Dhamdhere, D. J. Smith, M. Tsoi, and A. A. Demkov, Phys. Rev. B 87, 144422 (2013).
%Oxygen vacancy-mediated room-temperature ferromagnetism in insulating cobalt-substituted SrTiO3 epitaxially integrated with silicon

\bibitem{vasp96prb} G. Kresse and J. Furthmuller, Phys. Rev. B 54, 11169 (1996).
%Efficient iterative schemes for ab initio total-energy calculations using a plane-wave basis set  

\bibitem{vaspbackground} G. Makov and M. C. Payne, Phys. Rev. B 51, 4014 (1995).
%Periodic boundary conditions in ab initio calculations

\bibitem{HSE061}  J. Heyd, G. E. Scuseria, and M. Ernzerhof, The Journal of Chemical Physics 124, 219906 (2006).
%Erratum: Hybrid functionals based on a screened Coulomb potential J. Chem. Phys. 118, 8207

\bibitem{dudarev} Dudarev, S. L., Botton, G. A., Savrasov, S. Y., Humphreys, C. J., Sutton, Physical Review B, 57(3), 1505 (1998). 
%Electron-energy-loss spectra and the structural stability of nickel oxide: An LSDA+ U study.


\bibitem{pymatgen} S. P. Ong, W. D. Richards, A. Jain, G. Hautier, M. Kocher, S. Cholia, D. Gunter, V. L. Chevrier, K. A. Persson, and G. Ceder, Computational Materials Science 68, 314 (2013).
%Python Materials Genomics (pymatgen): A robust, open-source python library for materials analysis


\bibitem{matproj} A. Jain, S. P. Ong, G. Hautier, W. Chen, W. D. Richards, S. Dacek, S. Cholia, D. Gunter, D. Skinner, G. Ceder, and K. a. Persson, APL Materials 1, 011002 (2013).
%Commentary: The Materials Project: A materials genome approach to accelerating materials innovation


\bibitem{FINDSYM} Stokes, H. T., Hatch, D. M., Journal of Applied Crystallography, 38(1), 237-238, (2005). 
%FINDSYM: program for identifying the space-group symmetry of a crystal.

\bibitem{miguel2} Reyntjens, Peter D., et al., Materials 14.15 (2021).
%Ab-Initio Study of Magnetically Intercalated Platinum Diselenide: The Impact of Platinum Vacancies.

\bibitem{miguel1} Zhurkin, E. E., et al., Nuclear Instruments and Methods in Physics Research Section B: Beam Interactions with Materials and Atoms 269.14 (2011).

\bibitem{miguel3} Shiiba, Hiromasa, et al., Physical Chemistry Chemical Physics 15.25 (2013).
%Calculation of arrangement of oxygen ions and vacancies in double perovskite GdBaCo 2 O 5+ δ by first-principles DFT with Monte Carlo simulations.

\bibitem{materialsproject2} Hong Ding, Shyam S. Dwaraknath, Lauren Garten, Paul Ndione, David Ginley, and Kristin A. Persson
ACS Applied Materials \& Interfaces, 8 (20), 13086-13093  2016.
%the materials project 2

\bibitem{materialsproject3} Horton, M.K., Montoya, J.H., Liu, M. et al., npj Comput Mater 5, 64 (2019).
%the materials project 3

\bibitem{Tuller} Kuhn, M., Kim, J. J., Bishop, S. R. and Tuller, H. L., Chemistry Of Materials 25, 2970–2975 (2013).
%Oxygen Nonstoichiometry and Defect Chemistry of Perovskite-Structured Ba xSr-1-xTi 1– yFe yO 3-y/2+dSolid Solutions.

\bibitem{jmflorezmagnetite} J.M. Florez, J. Mazo-Zuluaga, and J. Restrepo. Hyperfine Interact, 161(1), 161-169 (2005).

\bibitem{moreno} Aramburu, J. A. and Moreno, M., J Phys Chem 125, 2284–2293 (2021).
%Key Role of Deep Orbitals in the dx2-y2-d3z2-r2 Gap in Tetragonal Complexes and 10Dq.

\bibitem{APS}  J. M. Florez, M. C. Onbasli, D. H. Kim, S. P. Ong, G. Ceder, P. Vargas, and C. A. Ross, Abstract: M32.00014, APS March Meeting 2015, Volume 60, Number 1 (2015).

\bibitem{SCO_OV} Tahini, H.A.; Tan, X.; Schwingenschlögl, U.; Smith, S.C.  ACS Catal. 2016, 6, 5565
%Formation and Migration of Oxygen Vacancies in SrCoO3 and Their Effect on Oxygen Evolution Reactions.

\bibitem{ovito} Alexander Stukowski Modelling Simul. Mater. Sci. Eng. 18 015012 (2010).


\bibitem{selfregu_changeoxida_insulator}  H. Raebiger, S. Lany, and A. Zunger, Nature 453, 763 (2008).
%Charge self-regulation upon changing the oxidation state of transition metals in insulators

\bibitem{chargedensi_vs_oxida_states} G. M. Dalpian, Q. Liu, C. C. Stoumpos, A. P. Douvalis, M. Balasubramanian, M. G. Kanatzidis, and A. Zunger, Phys. Rev. Materials 1, 025401 (2017).
%Changes in charge density vs changes in formal oxidation states: The case of Sn halide perovskites and their ordered vacancy analogues

\bibitem{Varignon:2017is} Varignon, J., Grisolia, M. N., Iniguez, J., Barthelemy, A., and Bibes, M. (2017). Npj Quantum Materials, 2(1), 21.
%Complete phase diagram of rare-earth nickelates from first-principles. 

%\bibitem{stfcpolaization} Cortés Estay, E. A., Ong, S. P., Ross, C. A.,  Florez, J. M.,  Magnetochemistry, 8(11), 144, (2022).
%Oxygen Deficiency and Migration-Mediated Electric Polarization in Magnetic Fe, Co-Substituted SrTiO3− δ.
 
\bibitem{Yan:2013ja} Yan, B., Jansen, M., and Felser, C., Nature Physics, 9(11), 709 (2013).
%A large-energy-gap oxide topological insulator based on the superconductor BaBiO<sub>3</sub>.


\bibitem{Kim:2018kf} Kim, N., Perry, N. H.,  Ertekin, E., Chemistry of Materials, 31(1), 233–243(2018).
%Atomic Modeling and Electronic Structure of Mixed Ionic–Electronic Conductor SrTi1–xFexO3–x/2+δ Considered as a Mixture of SrTiO3 and Sr2Fe2O5

\bibitem{BinOuyang:2019ep} Bin Ouyang, Chakraborty, T., Kim, N., Perry, N. H., Mueller, T., Aluru, N. R.,  Ertekin, E., Chemistry of Materials, (2019).
%Cluster Expansion Framework for the Sr(Ti1–xFex)O3–x/2 (0 < x < 1) Mixed Ionic Electronic Conductor: Properties Based on Realistic Configurations


\bibitem{ferro1} Xu, T., Shimada, T., Araki, Y., Wang, J. \& Kitamura, T, Nano Letters 16, 454–458 (2015). 
% Multiferroic Domain Walls in Ferroelectric PbTiO3 with Oxygen Deficiency.

\bibitem{ferro2} Lee, Kyoungjun, et al., Scientific reports 11.1,  1-9 (2021).
%Enhanced ferroelectric switching speed of Si-doped HfO2 thin film tailored by oxygen deficiency.

\bibitem{ferro3} Cheng, S. et al, Adv Funct Mater 26, 3589–3598 (2016). 
% Manipulation of Magnetic Properties by Oxygen Vacancies in Multiferroic YMnO3.

\bibitem{allenOccupationMatrixControl2014}  Allen, Jeremy P., and Graeme W. Watson., Physical Chemistry Chemical Physics 16.39, 21016-21031, (2014).

\end{thebibliography}

\end{document}